\documentclass[preprint, review, 10pt]{elsarticle}

\usepackage{graphicx}
\usepackage{fullpage}
\usepackage{amssymb}
\usepackage[normalem]{ulem}

\usepackage{lineno}
\usepackage{verbatim}
\usepackage{xcolor}
\usepackage{eqnarray}
\usepackage{siunitx}
\RequirePackage{ifthen}
\RequirePackage{graphicx}
\RequirePackage{amsmath}
\RequirePackage{fancyhdr}

\usepackage{fancyhdr,graphicx,amsmath,amssymb,microtype}
\usepackage[ruled,vlined]{algorithm2e}
\SetAlFnt{\small}
\SetAlCapFnt{\normalsize}
\SetAlCapNameFnt{\normalsize}
\usepackage{algorithmic}
\algsetup{linenosize=\small}
\include{pythonlisting}
\usepackage{setspace}

\usepackage[utf8]{inputenc}
\usepackage[english]{babel}

\definecolor{purp}{rgb}{0.4,0.2,0.8}

\usepackage{hyperref}
\definecolor{custom-blue}{RGB}{3,69,173}
\hypersetup{
    colorlinks = true,
    urlcolor   = blue,
    citecolor  = custom-blue,
}

\usepackage{caption}
\usepackage{subcaption}

\journal{Acta Materialia}

\begin{document}

\begin{frontmatter}


\title{Manifold learning for coarse-graining atomistic simulations: Application to amorphous solids}


\author[label1]{Katiana Kontolati}
\author[label1]{Darius Alix-Williams}
\author[label2]{Nicholas M. Boffi}
\author[label1]{Michael L. Falk}
\author[label2,label3]{Chris H. Rycroft}
\author[label1]{Michael D. Shields$^*$}

\address[label1]{Whiting School of Engineering, Johns Hopkins University, Baltimore MD 21218, USA}
\address[label2]{John A. Paulson School of Engineering and Applied Sciences, Harvard University, Cambridge, MA 02138, USA}
\address[label3]{Computational Research Division, Lawrence Berkeley National Laboratory, Berkeley, CA 94270, USA}

\begin{abstract}

  We introduce a generalized machine learning framework to probabilistically parameterize upper-scale models in the form of nonlinear PDEs consistent with a continuum theory, based on coarse-grained atomistic simulation data of mechanical deformation and flow processes. The proposed framework utilizes a hypothesized coarse-graining methodology with manifold learning and surrogate-based optimization techniques. Coarse-grained high-dimensional data describing quantities of interest of the multiscale models are projected onto a nonlinear manifold whose geometric and topological structure is exploited for measuring behavioral discrepancies in the form of manifold distances. A surrogate model is constructed using Gaussian process regression to identify a mapping between stochastic parameters and distances. Derivative-free optimization is employed to adaptively identify a unique set of parameters of the upper-scale model capable of rapidly reproducing the system's behavior while maintaining consistency with coarse-grained atomic-level simulations. The proposed method is applied to learn the parameters of the shear transformation zone (STZ) theory of plasticity that describes plastic deformation in amorphous solids as well as coarse-graining parameters needed to translate between atomistic and continuum representations. We show that the methodology is able to successfully link coarse-grained microscale simulations to macroscale observables and achieve a high-level of parity between the models across scales.

\end{abstract}

\begin{keyword}
Manifold learning \sep surrogate model \sep optimization \sep probabilistic learning \sep coarse-graining \sep parameter calibration \sep molecular dynamics simulation \sep amorphous solids \sep shear transformation zone


\end{keyword}

\end{frontmatter}




\section{Introduction}
\label{S:Background}


Physical phenomena often occur over a wide range of length and time scales; however, the observable scales of system models may be more restrictive. Scale limitations imposed by a single model can be rectified by combining models at different scales and passing information between them in a process known as \textit{multi-scale modeling}. This process leverages the accuracy of high-resolution models with the observation scales of lower-resolution models. For example, a multi-scale model of crack propagation in a simulated part may utilize molecular dynamics (MD) simulation near a crack tip and a continuum mechanics (CM) model elsewhere in a bulk specimen \cite{miller1998quasicontinuum}. MD simulations approximately model particle interactions at nanometer length and picosecond time scales and are often computationally intractable for systems larger than microns in size or for observation times longer than a microsecond. On the other hand, CM can model systems on human scales typical of many real systems but may require information from the lower-scale model to achieve a desired accuracy.

The process of transferring information from low to high scales is known as \textit{upscaling}. In this process, lower-dimensional models that function on coarser spatio-temporal scales (the so-called \textit{coarse-grained} (CG) models) are constructed via the process of \textit{coarse-graining}, during which the number of degrees-of-freedom (DOF) of a system is reduced, and often a continuum-level description is constructed which enables the use of larger simulation time-steps and spatial domains. The coarse-graining process generally involves a spatial and/or temporal averaging of atomic scale quantities, but the choice of length and time scales over which to perform averaging is not unique and introduces additional parameters into a model whose physical meaning may not be intuitive. A number of coarse-graining approaches exist; however, many are application-specific and a unified framework is altogether lacking. A tremendous body of literature exists on coarse-graining methodology. As written by Noid~\cite{noid2013perspective}, ``A vast array of approaches and philosophies have been adopted for achieving this goal. It seems quite challenging to catalog the resulting “landscape” of CG models in a completely coherent fashion.'' Recognizing this challenge, we attempt to provide an appropriate context for the proposed methodology by briefly summarizing some of the prevailing lines of research in coarse-graining methodology.

\subsection{Brief Review of Coarse-Graining Methodology}

Many coarse-graining methodologies take a ``bottom-up'' approach in which information from a lower-scale model is employed for the construction of an approximate potential of mean-force (PMF) for the CG model  \cite{akkermans2001structure, li2010characterizing, toth2007effective}. Essentially, the GC potential is determined as a systematic approximation to the complex many-body PMF, which for an all-atom (AA) system is able to describe all thermodynamic and structural properties that emerge on the length scales of the CG model. This class of methods includes the correlation function approaches such as the iterative Boltzmann inversion (IBI) \cite{reith2003deriving}, stochastic parametric optimization (SPO), reverse Monte Carlo (RMC) \cite{lyubartsev1995calculation,li2016comparative} and the conditional reverse work (CRW) \cite{brini2012chemically}. In these reverse-CG procedure approaches, macroscopic observables are reproduced by adjusting the parameters of the CG model to match specific probability distributions of the AA systems. A generalized technique called molecular renormalization group CG has been proposed, for the purposes of matching correlators obtained from atomistic and CG simulations, for observables that explicitly enter the CG Hamiltonian \cite{savelyev2009molecular}. An alternative class of bottom-up CG methods are the variational approaches, including entropy-based schemes for learning CG model parameters. These methods argue that the CG model optimally represents the atomistic model when the relative entropy of the system is minimized \cite{chaimovich2011coarse,carmichael2012new}. Another established variational method for determining optimal CG potentials is the multi-scale coarse-graining (MS-CG) method \cite{izvekov2005multiscale, izvekov2005systematic, villet2010numerical}, wherein optimization is employed to minimize a force-matching functional of the trial CG force-field. In contrast to other correlation function and variational approaches the MS-GC methods is not guaranteed to accurately reproduce any particular atomistic distribution function.

An alternative approach focuses on the calibration of atomistic reactive force-fields necessary to capture phenomenologically equivalent physical processes at intermediate spatio-temporal scales \cite{marrink2007martini}. These methods aim to handle a broader range of applications without the need to reparametrize the CG model. Such models have recently become popular for the study of soft matter (polymers and  biomacromolecules) \cite{gooneie2017review}. An extensive literature exists for the development of CG models for the study of biomolecular systems, for example \cite{noid2013perspective}. This approach has been applied for wider-ranging material classes as well. For instance, a strain energy conservation approach has been employed to calibrate parameterized force-fields to study large deformation and fracture of materials such as graphene \cite{ruiz2015coarse}. Representative volume element (RVE)-based approaches have also been employed to describe plasticity in amorphous solids \cite{homer2009mesoscale}. The limitation of these approaches is that a well-established procedure for the calculation of RVEs and a direct connection to thermodynamic principles are lacking.  Moreover, in a recent work, a methodology has been proposed that attempts to coarse grain atomistic data to parameterize thermodynamically consistent dynamical equations to model plasticity in amorphous solids (specifically metallic glasses) \cite{hinkle2017coarse}. The key feature of this methodology is that, rather than operating phenomenologically to empirically match observed dynamics of a physical system, it aims to operate entirely within a thermodynamically consistent theoretical framework and learn the essential physical constants from coarse-grained atomistic data. We adopt this coarse-graining framework herein.

In recent years, machine learning (ML) and data-driven methods have been successfully used to treat up-scaling problems. Many studies, for example, have attempted to accurately represent free-energy functions in the space of the atomistic degrees of freedom in an automated way without the need for substantial physical intuition and/or \textit{ad hoc} approximations \cite{chmiela2017machine,han2017deep,zhang2018deep}. In a recent study, a deep neural network (DNN)-based methodology called deep coarse-grained potential (DeePCG) aims to optimize the parameters of a many-body coarse-grained potential represented by a NN \cite{zhang2018deepcg}. In another recent study, the authors propose a deep learning-based coarse-graining framework which integrates big data of MD trajectories to optimize the parameters of an interatomic potential \cite{moradzadeh2019transfer}. A novel strategy based on generative adversarial networks (GANs) for CG parameterization called adversarial-residual-coarse-graining (ARCG) has been proposed, aiming to connect GANs-type \cite{goodfellow2014generative} implicit generative models with molecular models by optimizing traditional CG force-fields \cite{durumeric2019adversarial}. Moreover, several Bayesian methods have been successfully developed for the calibration and validation of CG models of atomistic systems in thermodynamic equilibrium as well as for uncertainty quantification for force field parameters \cite{farrell2014calibration,liu2008bayesian,schoberl2017predictive,angelikopoulos2012bayesian}. The Bayesian approach can also be used to treat problems characterized by high-dimensional inputs and a small number of training data obtained from fine-grained models and used for the generation of surrogate models \cite{rixner2020probabilistic} (meta-models or emulators) that enable rapid material behavior predictions based purely on past data. In a recent study, the authors proposed a hierarchical multiscale modeling technique in which a material failure classification model and stress regression models are trained from a dataset generated by an MD simulation, and are then implemented in a continuum model to reproduce material behavior at the macroscale \cite{xiao2019machine}. In \cite{wang2019machine}, a statistical learning-based technique was used to decompose the CG error and cross-validation to select and compare the performance of different CG models. A manifold learning-based approach using kernel PCA and diffusion maps to construct Gaussian process emulators for very high-dimensional output spaces arising from PDE model simulations was developed by Xing et al.~\cite{xing2016manifold}. Genetic algorithms (GA) and general polynomial chaos expansion (gPCE) based techniques have similarly been proposed for constructing coarse-scale representations of fine-scale models by reducing the dimensionality uncertainties propagated across scales \cite{gorguluarslan2014simulation}. An approach based on Gaussian process (GP) regression and a nonlinear autoregressive scheme \cite{kennedy2000predicting} capable of learning complex nonlinear and space-dependent cross-correlations between multi-fidelity models is developed by Perdikaris et al.~\cite{perdikaris2017nonlinear}. Finally, in another study CG PDEs are directly constructed from fine-scale data with the use of GP regression and neural networks (NN) \cite{lee2020coarse}.

\subsection{Proposed Coarse-Graining Approach}
\label{CG-intro}

Existing methodologies on the CG of detailed AA models have focused on the parameterization of approximate potentials and interactions that often sacrifice detail for higher computing efficiency. The well-established bottom-up approaches have not yet achieved their full modeling potential and so approximate CG models are often unable to capture the key underlying physics in cases of complex material systems. To tackle this issue, more complex terms can be added to the CG potential. However, the question of transferability arises. Moreover, many coarse-graining approaches take an \textit{ad hoc} approach wherein models are empirically or phenomenologically hypothesized, rather than being derived directly from first principles. Alternative approaches focus on the construction of ML-based models and cheap emulators that are often difficult to physically interpret since they usually lack important physical constraints and humanly interpretable parameters. Finally, the above approaches are often times application-specific and not easily transferable to different classes of materials.

To address the above challenges we propose a general ML-based optimization method called Grassmannian Efficient Global Optimization (Grassmannian EGO), which uses high-fidelity atomistic simulation data of material systems together with theoretical models in the form of PDEs to directly estimate the parameters of the latter given a choice of CG length scale and a desired low-to-high-level model accuracy. Rather than relying on phenomenological models, the proposed methdology considers models that are consistent with a continuum theory and incorporates data into that theory from lower-scale simulations. In particular, the proposed methodology is composed of four essential ingredients:
\begin{enumerate}
    \item A high-fidelity lower-scale model, e.g. MD, capable of simulating time evolution of the model material;
    \item A physically consistent continuum theory by which to develop evolution equations;
    \item A numerical solver capable of simulating time evolution the continuum system that is consistent with time evolution of the material produced by the lower-scale model;
    \item A \emph{hypothesized} means of transferring coarse-grained information from the lower-scale model to the continuum model.
\end{enumerate}
A few notes on these items follow. The approach is designed to be as general as possible. Hence, time evolution of the model material may represent many different physical processes, e.g. deformation, chemical reaction, heat transfer, etc. Hence, the continuum theory should be provided in an appropriate set of PDEs describing the physics of interest. In our application, we are specifically interested in inelastic mechanical deformation of a model material, but the framework is suitable for coarse-graining of nearly any physical process. In terms of the numerical solver, consistency is established by modeling similar time scales, length scales, and boundary conditions; albeit on far fewer degrees of freedom. Finally, a major challenge of coarse-graining is establishing the means of transferring information between models. Here, we enforce only that a means of this translation be hypothesized. Given this hypothesis, the proposed approach will generate a best fit continuum model. This does not ensure the validity of the hypothesis, but does provide a rigorous framework by which to conduct hypothesis testing and validation -- allowing the exploration of competing means of translating information, including those that incorporate aspects of machine learning.

The Grassmannian EGO adopts a manifold-projection based approach in which field solutions obtained from the CG fine-scale discrete model and a small number of forward continuum model evaluations are projected onto the lower-dimensional Grassmann manifold; a Riemannian topological space whose structure is exploited for measuring behavioral discrepancies between the models. A small training dataset is constructed and Gaussian process (GP) regression is employed to quantify the complex relation between the model and coarse-graining parameters and the solution error. Approximated solution error in terms of a manifold distance is represented by a high-dimensional objective (loss) function and is minimized using the EGO algorithm. Identification of an optimal set of inputs leads to calibration of the inexpensive PDE model capable of best reproducing material behavior with atomic-level detail. Finally, the proposed data-based methodology can be employed for the study of different classes of materials undergoing different physics, to investigate competing information translation hypotheses, and can be extended to investigate parameter sensitivity related to material processing and compositional variations for example.

\subsection{Multi-scale modeling of amorphous solids: Coarse-graining metallic glasses}

Amorphous solids are a vast and important class of materials that possess unique and challenging properties that make the physics of their formation and the mechanics of their deformation especially difficult to understand and model. In this work, we specifically study metallic glasses (MGs) as an exemplar for the broader class of amorphous solids. MGs are formed when liquid metals are rapidly cooled such that the liquid bypasses crystallization and undergoes a glass transition in which the microstructure remains in its liquid-like disordered configuration, rather than undergoing first-order phase transition \cite{wang2004bulk}. Moreover, MGs possess favorable properties, such as high yield strength and toughness \cite{das2005work, schroers2004ductile, schuh2007mechanical, trexler2010mechanical, hufnagel2016deformation}.

One of the main challenges in multi-scale modeling of amorphous solids is that, unlike crystalline materials where macroscale deformation mechanisms can be directly linked to well-known point defects in the microstructure, the notion of a defect in an amorphous material is not well-established. This makes multi-scale modeling, and in particular coarse-graining, quite difficult because properties such as defect densities that are naturally used to span length-scales are difficult to define. This has led to a proliferation of phenomenological models that are capable of modeling certain aspects of MG deformation (e.g.\ \cite{anand2007constitutive, demetriou2009coarse}), but lack a direct connection to underlying microstructural changes that occur during inelastic deformation.



In this work, we leverage the proposed methodology to model plasticity in a metallic glass alloy by coarse-graining atomistic simulations of shear deformation to inform a continuum model \cite{rycroft2015eulerian,boffi20projection} of the shear transformation zone (STZ) theory of plasticity \cite{bouchbinder2009nonequilibrium}. Building on previous coarse-graining efforts of the authors \cite{hinkle2017coarse}, the proposed methodology provides a machine learning formalism by which to directly coarse-grain atomistic data to inform the STZ theory. STZ theory is a thermodynamically grounded theory that presumes that deformation in a glass can be described by a mean-field of point defects called STZs that mediate shear deformation, which have been directly observed in atomistic simulations \cite{manning2011vibrational, patinet2016connecting,cubuk2017structure,richard2020predicting}. It further presumes that the densities of these defects can be translated to higher length-scales through an effective ``disorder'' temperature \cite{langer2004dynamics, shi2007evaluation, bouchbinder2009nonequilibrium, falk2011deformation}. Codes and data for reproducing the results in this paper can be found at: https://github.com/katiana22/GrassmannianEGO.

\section{Manifold Learning Principles}

In this section, we review the relevant components of the manifold learning methods employed for the proposed surrogate-based coarse-graining framework of Section \ref{Grassmannian_EGO}. First, we introduce the Grassmann manifold and Grassmannian projection, which is employed to reduce field quantities derived from multiscale simulations into points on a manifold, such that distances between the fields can be measured. These distances function to assess the similarity of particular micro-scale (e.g. coarse-grained molecular dynamics) and macro-scale (e.g. continuum mechanics) model solutions. We then present the necessary components of Gaussian process (GP) regression, a supervised learning method used to construct surrogate models endowed with a Gaussian uncertainty measure. In the proposed methodology, this surrogate model is used to fit data that correspond to the distance (again, similarity) between models at different scales; effectively serving as an approximation of the objective function for parameter optimization. Finally, the Efficient Global Optimization (EGO) framework is introduced. EGO exploits the GP surrogate and its uncertainty measure in order to efficiently identify important surrogate model training points that are used to determine a global minimum. Subsequently, the EGO is used to identify parameters that result in the minimum Grassmannian distance between models at different scales.

\subsection{Grassmann Manifold Projection}
\label{grass-proj}

Data generated by physics-based models are typically described in high-dimensional spaces; however due to physical constraints they usually live on lower-dimensional subspaces. Performing tasks such as measuring the distance between solutions or predicting similar ones (via interpolation) in the original space may lead to systematic errors due to the existence of noisy and unimportant features (or dimensions). Such errors can be avoided via linear or nonlinear dimension reduction. Linear dimension reduction techniques are often preferred as they are computationally cheap compared to nonlinear methods since they only require simple matrix multiplications. Orthogonal projections specifically, an approach that has been widely used in computer vision for images and videos \cite{turaga2008statistical}, can be employed for reduction of full solutions by projecting them onto a compact lower-dimensional encoding or manifold whose structure can be then utilized and explored. Such projections are able to preserve essential information and complexity within a more compact representation \cite{breger2020orthogonal}. Grassmann manifolds, in particular, are well suited for such applications since they possess a smooth differentiable topological structure, are conveniently derived from large datasets of arbitrary shape, and possess a direct means of comparing the subspaces that compose the manifold \cite{absil2004riemannian}.



The Grassmann manifold $\mathcal{G}(p, n)$ with integers $n \ge p > 0$ is a topological space formed by all $p$-dimensional subspaces existing in an $n$-dimensional vector space $\mathbb{R}^{n}$. A point (subspace) $\mathbf{U}$ on the Grassmann manifold is  represented as an $n \times p$ ordered orthonormal matrix, i.e. $\mathbf{U}^{\top} \mathbf{U} = \mathbf{I}_p$, where $\mathbf{I}_p$ is the identity matrix. Thus, the Grassmann manifold is the collection of all subspaces
\begin{linenomath*}
    \begin{equation}
    \mathcal{G}(p, n) = \{ \mbox{span}(\mathbf{U}) : \mathbf{U} \in \mathbb{R}^{n \times p} : \mathbf{U}^{\top} \mathbf{U} = \mathbf{I}_p \}.
    \end{equation}
\end{linenomath*}

The Grassmannian projection of a dataset requires as a first step the reshaping of each high-dimensional data set into a 2D matrix.
Projection of an arbitrary data point, defined through the matrix $\mathbf{X}\in \mathbb{R}^{n\times m}$, onto the Grassmann manifold is performed by decomposing the matrix as
\begin{linenomath*}
    \begin{equation}
        \mathbf{X} = \mathbf{U} \mathbf{\Sigma} {\mathbf{V}}^{\top}
    \end{equation}
\end{linenomath*}
using the thin Singular Value Decomposition (thin SVD), where (after a suitable truncation) the columns of the $n \times p$ matrix $\mathbf{U}$ and $m \times p$ matrix $\mathbf{V}$ contain orthonormal singular vectors such that $\mathbf{U}^{\top}\mathbf{U} = \mathbf{I}$ and $\mathbf{V}^{\top}\mathbf{V} = \mathbf{I}$, and $\mathbf{\Sigma}$ is a $p \times p$ diagonal matrix whose non-zero elements are the singular values ordered by magnitude. Thus, $\mathbf{U}\in\mathcal{G}(p, n)$ and $\mathbf{V}\in\mathcal{G}(p, m)$.



Next, let $\mathbf{A}$ and $\mathbf{B}$ represent arbitrary data sets. In our case these are field quantities obtained by two numerical models, a coarse-grained snapshot of a molecular dynamics simulation and a corresponding snapshot of an upper-scale continuum model informed by the coarse-grained atomistic data. Singular value decomposition of these fields yields $\mathbf{A} = \mathbf{U}_A \mathbf{\Sigma}_A {\mathbf{V}_A}^{\top}$ and $\mathbf{B} = \mathbf{U}_B \mathbf{\Sigma}_B {\mathbf{V}_B}^{\top}$. If $\mathbf{A}$ and $\mathbf{B}$ can be represented by elements on the Grassmann manifold $\mathbf{U}_A, \mathbf{U}_B \in \mathcal{G}(p, n)$, the distance between their representative subspaces can be measured using the principal angles between them $\{\theta_i\}_{i=1}^{p}$. When $\mathbf{U}_A$ and $\mathbf{U}_B$ are ordered by decreasing magnitude of their respective singular values, the singular values of ${\mathbf{U}_A}^{\top} \mathbf{U}_B$ are the cosines of the principal angles. That is, letting $\mathbf{C} \equiv {\mathbf{U}_A}^\top \mathbf{U}_B$, the principal angles are found by decomposing
\begin{equation}
    \mathbf{C} = \mathbf{U}_C \mathbf{\Sigma}_C {\mathbf{V}_C}^{\top},
\end{equation}
Denoting the diagonal entries of $\mathbf{\Sigma}_C$ by $\sigma_i$, the principal angles are computed as $\theta_i = \cos^{-1}\sigma_i$ for $i = 1, \ldots, p$. The principal angles are the smallest angles between all bases from both subspaces.




The distance measure, $d = d(\theta_i), i = 1, 2, \ldots, p$, provides a measure of the distance between two points $\mathbf{U}_A, \mathbf{U}_B \in \mathcal{G}(p, n)$. A number of different distance measures are defined using the principal angles \cite{zhang2018grassmannian}. Here, we use the Grassmann distance $d_{\mathcal{G}(p, n)}$ (also referred to as the geodesic distance or arc-length), which measures the geodesic. The geodesic is the shortest path between two points on the manifold $\mathcal{G}(p, n)$.  The Grassmann distance between points $\mathbf{U}_A$ and $\mathbf{U}_B$ is given by
\begin{equation}
\label{eq:Grassmann_Distance}
    d_{\mathcal{G}(p,n)}(\mathbf{U}_A, \mathbf{U}_B) = \bigg( \sum_{i=1}^{p} {\theta_i}^2 \bigg)^{1/2} = \left\| \cos^{-1} \mathbf{\Sigma}_{C} \right\|_F
\end{equation}
where subscript $F$ indicates the Frobenius norm.

Practically speaking, the distances computed in Eq.\ \eqref{eq:Grassmann_Distance} mean that we are comparing solutions based on their column spaces. We could likewise compute distances based on their row spaces using $\mathbf{V}_A$, $\mathbf{V}_B \in \mathcal{G}(p, m)$ or devise a composite distance that accounts for both differences in the column space and the row space. Moreover, it follows from the following equation
\begin{equation}
    c\mathbf{X} = \mathbf{U} (c\mathbf{\Sigma}) {\mathbf{V}}^{\top},
\label{properties}
\end{equation}
that the geodesic distance does not account for behavioural discrepancies related to scaling the data points. Instead, it measures differences related to the structural signature of the data points. To account for differences in the magnitude of the points, alternative distance metrics can be employed, for example that incorporate information from the singular values $\mathbf{\Sigma}$ resulting from the SVD.

Singular value decomposition of response fields from the respective models may yield eigenvalues with small values. If a threshold, e.g.\ $\sigma_\textit{cut}$, is used to determine the truncation of $\mathbf{U}_A$ and $\mathbf{U}_B$, the following approximation matrices may result
\begin{equation}
\begin{aligned}
    \mathbf{U}_A \rightarrow \hat{\mathbf{U}}_A, \\
    \mathbf{U}_B \rightarrow \hat{\mathbf{U}}_B,
\end{aligned}
\end{equation}
where $\hat{\mathbf{U}}_A$ is $n \times r_A$ and $\hat{\mathbf{U}}_B$ is $n \times r_B$ and $r_A$ and $r_B$ are the respective ranks. If the ranks of the approximation matrices differ, i.e.\ $r_A \ne r_B$, their subspaces will exist on different manifolds $\mathcal{G}(r_A,n) \ne \mathcal{G}(r_B,n)$. To measure the distance between the approximation matrices we must first embed the subspaces in a doubly infinite Grassmannian manifold $\mathcal{G}(\infty, \infty)$ which can be viewed as the disjoint union of all $r_i$-dimensional subspaces over all $r_i \in \mathbb{N}$ as
\begin{equation}
    \mathcal{G}(\infty, \infty) = \coprod^{\infty}_{r_i = 1} \mathcal{G}(r_i, \infty).
\end{equation}
The modified distance can be computed as
\begin{equation}
    d_{\mathcal{G}(\infty,\infty)}(\mathbf{U}_A, \mathbf{U}_B) = \bigg( \sum_{i=1}^{\min(r_A, r_B)} {\theta_i}^2  + |r_A - r_B| \pi^2/4 \bigg)^{1/2}.
\end{equation}
 A second term is introduced to Eq.~\eqref{eq:Grassmann_Distance} to account for missing eigenvectors in the lower dimensional subspace by assuming that they are completely orthogonal to those of the higher dimensional subspace. The interested reader is referred to \cite{ye2016schubert} for more information on this procedure.

We now have a means of compactly representing complex response fields of multi-scale models as well as a metric to quantify the discrepancy between these representations.

\subsection{Gaussian Process Regression} \label{GPR}
Gaussian process (GP) regression is a widely used supervised learning method used for constructing surrogate models with uncertainty measurements derived from small datasets. Standard regression models often assume that errors are independent, however this assumption is usually false since, for a deterministic model, poor fitting is due entirely to modeling error (inefficient set of regression terms) and not noise or measurement error \cite{jones1998efficient}. The GP
model, in contrast, assumes that errors are correlated and, in the high-dimensional function space, take the form of a Gaussian stochastic process as
\begin{equation}
    f(\mathbf{x}) \sim \mathcal{GP}(m(\mathbf{x}),k(\mathbf{x},\mathbf{x'}))
\label{surr}
\end{equation}
where $m(\mathbf{x})$ is the mean function and $k(\mathbf{x},\mathbf{x'})$ is the covariance function.

A common covariance function is the radial basis function (RBF) or squared exponential (SE). This well-behaved infinitely differentiable positive definite function specifies the covariance between pairs of random variables as
\begin{equation}
    \mathrm{cov}(y_p,y_q) =   k(\mathbf{x_p},\mathbf{x_q}) + \sigma_n^2 \delta_{ij} =  \exp{(-\frac{1}{2}|\mathbf{x_p}-\mathbf{x_q}|^2}) + \sigma_n^2 \delta_{ij}
\label{cov}
\end{equation}
where $\mathbf{x_p}$ and $\mathbf{x_q}$ are two input random variables, $\sigma_n^2$ is the variance of the i.i.d. Gaussian noise $\epsilon$ and $\delta_{ij}$ is the Kronecker delta.

Let $X$ correspond to training input points and $X_*$ to test input points. If we calculate elementwise the covariance function using Eq.~\eqref{cov} we can express the model training output as a Gaussian random vector as
\begin{equation}
    y \sim \mathcal{N}(0,K(X,X) + \sigma_n^2 I)
\label{g_vector}
\end{equation}
where $I$ is the identity matrix. Note that location invariant zero-mean GP modeling is often the default in the machine learning and surrogate modeling literature and, perhaps surprisingly, this special case works quite well \cite{williams2006gaussian}.

The joint distribution of the training outputs, $y$, and the test outputs $\mathbf{f_*}$ according to the
prior is
\begin{equation}\label{joint_dist}
\begin{bmatrix}
    y \\
    \mathbf{f_*}
\end{bmatrix}
\sim \mathcal{N}
\begin{pmatrix}
    0, \begin{bmatrix}
    K(X,X) + \sigma_n^2 I, \ \ \ K(X,X_*) \\
    K(X_*,X) \ \ \ \ \ \ \  K(X_*,X_*)
\end{bmatrix}
\end{pmatrix}.
\end{equation}
With $n$ training points and $n_*$ testing points, $K(X,X_*)$ denotes the
$n \times n_*$ matrix of the covariances evaluated at all pairs of training and test points, and likewise for the other matrices $K(X, X)$, $K(X_*, X_*)$ and $K(X_*, X)$.

The last step is to condition the joint Gaussian prior distribution on the observations in order to derive the key predictive equations as
\begin{equation}
    \mathbf{f_*}|X,y,X_* \sim \mathcal{N}(\mathbf{\bar{f}_*},\mathrm{cov}(\mathbf{f_*}))
\label{conditional}
\end{equation}
where
\begin{equation}
    \mathbf{\bar{f}_*} \triangleq \mathbb{E}[\mathbf{f_*}|X,y,X_*] = K(X_*,X)[K(X,X) + \sigma_n^2 I]^{-1}y,
\label{values_f}
\end{equation}
\begin{equation}
    \mathrm{cov}(\mathbf{f_*}) = K(X_*,X_*) - K(X_*,X)[K(X,X) + \sigma_n^2 I]^{-1} K(X,X_*).
\label{values_f2}
\end{equation}
As a practical note, the covariance function typically possesses a set of hyperparameters. For example, the RBF covariance function takes the following form
\begin{equation}
    k(\mathbf{x}_p,\mathbf{x}_q) = \sigma_f^2 \exp{(-\frac{1}{2 l^2}(\mathbf{x}_p-\mathbf{x}_q)^2)} + \sigma_n^2 \delta_{ij}
\label{cov2}
\end{equation}
where $\sigma_f^2$ is the signal variance and $l$ the characteristic length scale. The variance $\sigma_f^2$ is a scaling parameter that controls the magnitude of the predictive uncertainty depends on the amplitude of the training data points. The length scale $l$, is the rate of decay of correlation and controls the smoothness of $k$ by setting the distance over which one expects points to be correlated.

For multidimensional GPs, we express the signal variance $\sigma_f^2 \rightarrow \sigma_{fp}^2$, the length scale $l \rightarrow l_p$ and the noise variance $\sigma_n^2 \rightarrow \sigma_{np}^2$, to highlight that these parameters are taken with respect to a given coordinate direction $p$. Hyperparameter tuning must be performed in order for the model to understand the typical behavior of the observed data. Typically, likelihood-based inferential schemes are utilized due to their easy automation and generalization to higher dimensional hyperparameter spaces, but other methods are also used such as cross validation (CV). Maximum-likelihood estimates require the maximization of the log-likelihood and since closed-form solutions are not always possible, numerical optimization methods may be employed.

Finally, standardization of the data is desirable in the case where a zero-mean GP model is chosen. Numerical issues can also be remedied by this practice, as scaling the data alleviates the issue of converting an ill-conditioned covariance matrix. The reader is referred to \cite{rasmussen2003gaussian,gramacy2020surrogates} for more details.

\subsection{Efficient Global Optimization}

The most straightforward way to use surrogate  models for optimization is to fit a surface to the objective function and search for the minimum of the surface. This simple process however, is prone to errors from the surrogate model and is especially susceptible to local minima. The EGO framework leverages the GP surrogate model and its Gaussian uncertainty measure to iteratively resolve the objective function in areas where minima are likely. In particular, the surrogate is initially trained at an initial set of points, often generated using a space-filling design such as Latin hypercube sampling or a Sobol sequence. Knowledge of the expected values and uncertainties in the objective function from the GP is then used to decide where future sampling is performed in order to improve the model and identify the global minimum.

To decide the location of each new training point, a learning function known as the expected improvement (EI) function is used \cite{jones1998efficient}. The EI balances the trade-off between locally exploring existing minima and broadly exploring the function in order to avoid getting stuck in local minima. More specifically, the improvement of the surrogate model at point $j$ is given by
\begin{equation}
    I^{(j)} = \max{(f_\textit{min}-Y^{(j)},0)}
\end{equation}
where $f_\textit{min}$ is the current minimum of the function
\begin{equation}
    f_\textit{min} = \min{(y^{(1)}, \ldots, y^{(n)}})
\end{equation}
and
\begin{equation}
    Y^{(j)} \stackrel{iid}{\sim} \mathcal{N}(\hat{y}^{(j)},s^{(j)})
\end{equation}
where $\hat{y}$ is the prediction of the surrogate model and $s$ its standard error at $\mathbf{x}$. This latter term defines the uncertainty in the prediction of the surrogate model.

Taking the expectation of this improvement yields \cite{jones1998efficient}
\begin{equation}
    E[I(\mathbf{x}^{(j)})] = (f_\textit{min} - \hat{y}^{(j)}) \Phi\bigg(\frac{f_\textit{min} - \hat{y}^{(j)}}{s^{(j)}}\bigg) + s^{(j)} \phi\bigg(\frac{f_\textit{min} - \hat{y}^{(j)}}{s^{(j)}} \bigg)
\label{EI}
\end{equation}
where $\Phi(\cdot)$ is the standard normal cumulative distribution function and $\phi(\cdot)$ is the standard normal probability density function. The EI will be exactly zero if $\hat{y}^{(j)} = \hat{y}^{(i)}$, i.e.\ evaluation is performed at a sample point from the training dataset $\mathcal{D}$, and positive elsewhere. Given $n$ initial observations (evaluations of the objective function), point $(n + 1)$ is chosen to maximize the EI, i.e.\ $\max(E[I(\mathbf{x})])$.


The EI function is highly multi-modal, but it is expressed in closed form and monotonically increases as $\hat{y}$ decreases and $s$ increases. This monotonicity is utilized when trying to find the maximum, using a branch-and-bound algorithm. The interested reader can find details in the work of Jones et al.~\cite{jones1998efficient}.


\section{Grassmannian EGO for coarse-graining} \label{Grassmannian_EGO}

\begin{figure}[ht!]
\begin{center}
\includegraphics[width=0.8\textwidth]{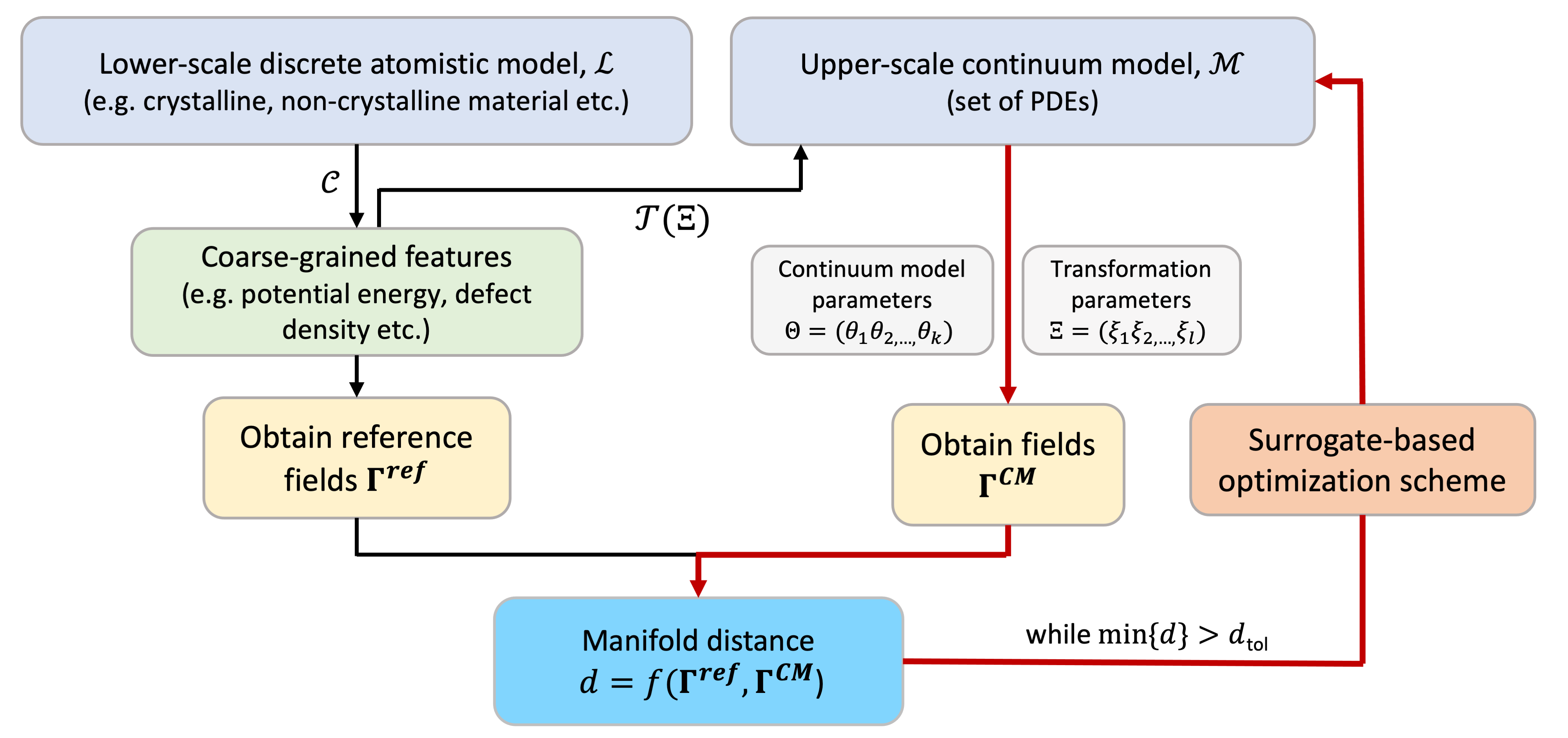}
\caption{A schematic illustrating Grassmannian EGO as a general methodology for coarse-graining atomistic simulations for various classes of materials.}
\label{fig:f1}
\end{center}
\end{figure}

In this section, we present a general Grassmannian learning methodology for coarse-graining lower-scale atomistic simulations to calibrate continuum level PDEs or other parameterized upper-scale models. The proposed framework combines the presented concepts of low-dimensional manifold learning and the EGO algorithm through the steps elucidated in detail in the sequel. A graphical illustration of the objective of Grassmannian learning as a general methodology for CG atomistic models is shown in Figure \ref{fig:f1}.

\subsection{Lower-Scale Modeling}

Consider a lower-scale material model ($\mathcal{L}$) capable of describing in detail the behavior of the microscopic particles within the material, e.g.\ a molecular dynamics (MD) model. The simulation of such models, especially when run on long timescales, is usually computationally expensive which precludes its use for large materials systems. This lower-scale model serves as the basis for coarse-graining and is assumed to be of sufficient detail so as to provide a sufficiently close approximation of ``ground truth'' material behavior.

Additionally, consider an operator $\mathcal{C}:\mathbb{R}^p\to \mathbb{R}^q$, $q\ll p$ that serves to reduce the number of degrees of freedom of this model, the so-called coarse-graining operator. This operator could be as simple as a moving average over atomic-scale quantities (such as atomic potential energies) or as complex as a differential operator. When assigning the operator $\mathcal{C}$, it is also generally necessary to select an appropriate coarse-graining length-scale (although certain classes of problems may have intrinsic physical length-scales that make this selection clear). In the proposed methodology, it is also possible to learn this coarse-graining length-scale directly if necessary.


From numerical simulations of $\mathcal{L}$, full-field solutions are obtained in the form of generalized tensors, which can be coarse-grained through $\mathcal{C}$ and arbitrarily reshaped.
A common practice is to transform the existing vectorized field quantities into  matrices $\mathcal{Z}_{i}^* \in \mathbb{R}^{n \times m}$ where $n \times m$ is the number of the degrees of freedom and $i=1,\ldots,N_{f}$ where $N_{f}$ is the total number of simulation frames, or snapshots. Although vectorized solutions can be reshaped arbitrarily, a general rule-of-thumb is to reshape each solution as close to a square matrix as possible. Reshaping of solutions does not affect the dependency between its individual components, as it simply requires a re-indexing, however consistent reshaping across all solution snapshots is essential. Generically, these fields in matrix form will be denoted by $\Gamma^{\textit{ref}}$ and will serve as the reference (or target) of the proposed optimization framework. As we discuss later, consistency between dimensions of the lower-scale atomistic coarse-grained solutions and continuum model solutions must be ensured, for comparison purposes.

\subsection{Training Simulations from Upper-Scale Model}
Consider next an upper-scale material model $\mathcal{M}$, which is based on a physically consistent continuum theory and is typically described by a set of partial differential equations (PDEs) having (potentially stochastic) parameters $\Theta=\{\theta_1, \ldots,\theta_k\}$ where $k$ is the number of parameters or features, and specified boundary conditions. When solved numerically, this model operates on fewer degrees of freedom than the lower-scale model $\mathcal{L}$, is much faster computationally, and is capable of simulation over a much larger range of spatio-temporal scales. Let us further recognize that the upper-scale model $\mathcal{M}$ may rely upon information passed (or mapped) directly from simulation of the lower scale model $\mathcal{L}$ through a transformation $\mathcal{T}:\mathbb{R}^l\to \mathbb{R}^l$ having parameter vector $\Xi=\{\xi_1, \dots, \xi_l\}$ referred to as the transformation parameter vector. The goal is to calibrate the combined parameter vector $X=\{\Theta, \Xi\}$ such that simulations of $\mathcal{M}$ reproduce the coarse-grained results of $\mathcal{L}$ as accurately as possible.


We start by generating $\mathcal{N}$ realizations (observations) $\{X_j\}_{j=1}^{\mathcal{N}}$, of the random vector $X = \{\theta_1, \ldots,\theta_k;\xi_1, \dots, \xi_l\}$ consisting of all $k+l$ uncertain parameters. We use a space-filling experimental design such as Latin hypercube sampling (LHS), Sobol sequence, etc. Determining a distribution and the bounds of the uncertain inputs may require a preliminary investigation of their sensitivity on the model $\mathcal{M}$. Alternatively, the distribution and bounds may be determined by physical intuition or constraints, i.e.\ certain parameter combinations may produce physically impossible conditions. A good practice is to sample the bounds of the parameter space in order avoid extrapolation errors in the surrogate models.

As in the lower-scale simulations, field solutions of the upper-scale model in the form of generalized tensors are reshaped into matrices $\mathcal{Z}_{i,j} \in \mathbb{R}^{n \times m}, \ i=1,\dots,N_f, j=1,\dots, \mathcal{N}$. Solution dimension is dictated by the numerical discretization and must be consistent with the lower-scale field solutions (i.e.  $\mathcal{Z}_{i}^*, \mathcal{Z}_{i,j} \in \mathbb{R}^{n \times m}$) such that the coarse-grained lower scale model and the upper-scale model have matching degrees of freedom.


\subsection{Low-Dimensional Manifold Projection}
Reference solutions in the form of matrices obtained by the lower-scale model ($\mathcal{Z}_{i}^*\in \mathbb{R}^{n \times m}$) and training solutions from the upper-scale model ($\mathcal{Z}_{i,j} \in \mathbb{R}^{n \times m}$), are now projected onto the Grassmann manifold by factoring into ordered orthonormal matrices using a thin SVD. For simplicity in notation, we will drop the indexing, $j$, on training realization and notate a single realization of the upper-scale solution. That is, we perform the following operations:
\begin{equation}
    \mathcal{Z}_{i}^* = \mathbf{U}_{i}^* \mathbf{\Sigma}_{i}^* {\mathbf{V}_{i}^*}^{\top} \ \ \ , \ \ \  \mathcal{Z}_{i} = \mathbf{U}_{i} \mathbf{\Sigma}_{i} {\mathbf{V}_{i}}^{\top}
\label{man_learn_1}
\end{equation}
where
\begin{equation}
    \mathbf{U}_{i}^*\in \mathbb{R}^{n \times p^*}, \ \ \mathbf{\Sigma}_{i}^*\in \mathbb{R}^{p^* \times p^*}, \ \ \mathbf{V}_{i}^*\in \mathbb{R}^{m \times p^*}
\label{man_learn_2}
\end{equation}
\begin{equation}
    \mathbf{U}_{i}\in \mathbb{R}^{n \times p}, \ \ \mathbf{\Sigma}_{i}\in \mathbb{R}^{p \times p}, \ \ \mathbf{V}_{i}\in \mathbb{R}^{m \times p}.
\label{man_learn_3}
\end{equation}

\begin{figure}[ht!]
\begin{center}
\includegraphics[width=0.7\textwidth]{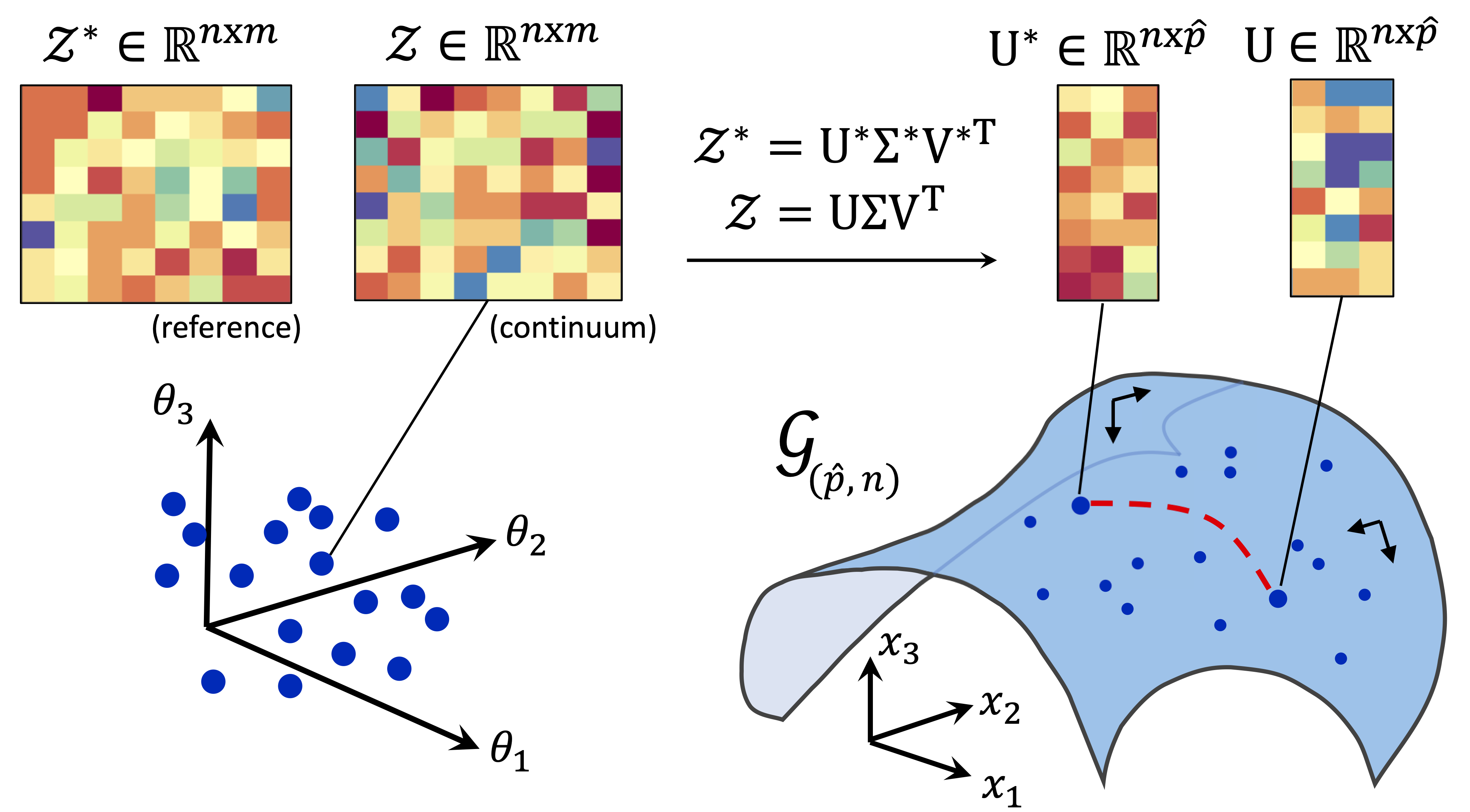}
\caption{A schematic of the low-dimensional manifold projection. Full-field solutions from the reference lower-scale model $\mathcal{L}$ and the observed upper-scale model $\mathcal{M}$, $\mathcal{Z}_{i}^*$ and $\mathcal{Z}_{i}$ respectively, are projected onto the Grassmann manifold $\mathcal{G}(\hat{p},n)$. The geodesic distance between the reduced representations ($U_{i}^*$,$U_{i}$) is used to measure the behavioural discrepancies between the multi-scale models.}
\label{fig:f2}
\end{center}
\end{figure}

Ranks obtained by the linear decomposition may not be equal, i.e. $p^* \ne p$ in general, and in such case the maximum between the two ranks is retained, i.e. $\widehat{p} = \max (p,p^*)$, so that no essential structural information is lost. Matrices $\mathcal{Z}_{i}^*$ and $\mathcal{Z}_{i}$ can be now represented as points or subspaces on the Grassmann manifold where $\mathbf{U}_{i}^*, \mathbf{U}_{i} \in \mathcal{G}(\widehat{p}, n)$. The geodesic manifold distance $d_{\mathcal{G}(\widehat{p}, n)}$ between the reduced representations of field quantities is then computed by Eq.~\eqref{eq:Grassmann_Distance} for each snapshot. A schematic illustrating the manifold-projection is shown in Figure \ref{fig:f2}.

\subsection{Define the objective function}

The objective of the coarse-graining is to minimize the difference between response of the lower-scale model and the upper-scale model. The geodesic distances defined in the previous section provide a means of comparing field response quantities for any given solution snapshot. For problems involving mechanical deformation as studied here, multiple snapshots of the physical field solutions that correspond to various states of the material (e.g.\ linear-elastic, yielding and plastic responses) are chosen so that the aggregate mechanical behavior is captured. Physical field quantities that may be used to compute distance metrics for comparison purposes are stress fields, strain fields, temperature fields, displacement fields, etc.

However, we may desire to endow our objective function with addition response measures such as more conventional scalar or time-history response measures. In other words, the geodesic distance is one measure of similarity of response quantities but numerous other measures may be incorporated in the assessment of discrepancy between lower-scale and upper-scale model response. For example, when studying mechanical deformation we may be interested in ensuring that the models produce consistent overall stress--strain response. When comparing stress--strain curves $(\tau-\gamma)$, for example, a conventional point-wise $L_2$ norm may provide an appropriate distance measure.

In the case where multiple distance measures exist, the overall distance (objective function) is computed through a weighted summation of the individual distances as:
\begin{equation}
\Delta = \sum_{i=1}^{N_d} w_i d_i
\label{eqn:d_total}
\end{equation}
where $w_i$ are normalization coefficients assigning weights to each of the $N_d$ distance measures, $d_i$ (Grassmannian or otherwise). Coefficients are chosen by considering the magnitudes of the calculated distance of each physical quantity and the relative importance given to each metric.


\subsection{EGO parameter optimization using Gaussian process surrogate objective function}

Distances are computed for each of the $\mathcal{N}$ training samples $X_i$. By appending the vector of distances $\mathbf{D}=\{\Delta_1, \dots, \Delta_{\mathcal{N}}\}$ to the set of initial observations $\mathbf{X}=\{X_1,\dots, X_\mathcal{N}\}$ we obtain the \textit{training dataset} $\mathcal{D} = \{X_i,\Delta_i\}_{i=1}^{\mathcal{N}}$. We will occasionally use the term \textit{labels} for the distances $\Delta_i$, \textit{samples} for the parameter realizations $\mathbf{X}$, and \textit{features} for the uncertain parameters $\Theta_i$, which is consistent with machine learning terminology. The above process is expected to be computational fast, since a relatively small number of inexpensive upper-scale model observations is required to form the training dataset $\mathcal{D}$.


GP regression is performed on the training dataset $\mathcal{D}$. The GP model,
\begin{equation}
    f(\mathbf{x}) \sim \mathcal{GP}(m(\mathbf{x}),k(\phi, \mathbf{x},\mathbf{x'})),
\end{equation}
\noindent
with mean function $m(\mathbf{x})$ and covariance function $k(\phi, \mathbf{x},\mathbf{x'})$ with hyperparameters $\phi$, aims to learn the underlying distribution of the objective function from the data using a Bayesian approach. The model is then used to draw samples and make predictions within the EGO algorithm.




Consider now the generation of $\mathcal{N}^*$ additional samples, where $\mathcal{N}^*\gg\mathcal{N}$, for which we want to predict the corresponding distances $\{\Delta^{*}_{i}\}_{i=1}^{\mathcal{N}^*}$ along with their associated uncertainty. We refer to the additional samples as the set of \textit{test points} $\mathbf{X}^*$. The covariance matrices at all pairs of training $X$ and test points $X^*$, are evaluated to form the joint prior distribution as in Eq.\ \eqref{joint_dist}. Finally the conditional distribution
\begin{equation}
    \mathbf{f_*}|X,y,X^* \sim \mathcal{N}(\mathbf{\bar{f}_*},\mathrm{cov}(\mathbf{f_*}))
\end{equation}
is calculated with a predictive mean and covariance function as in Eqs.\ \eqref{values_f}~\&~Eq.\ \eqref{values_f2} respectively. Each predicted point is an approximation of the solution error, a measure of similarity between the upper-scale model and the lower-scale reference model solutions at a point in the parameter space. This error is treated as a random variable with mean value and standard deviation dictated by the statistics of the above calculated conditional distribution.

A very large number of computationally inexpensive surrogate evaluations is performed using Monte Carlo simulation (MCS) and the EI function (Eq.\ \eqref{EI}) is computed at each point, where $f_\textit{min} = \min \mathbf{\{\mathbf{y}}_i\}_{i=1}^{\mathcal{N}}$ which is simply the smallest distance obtained by the training dataset $\mathcal{D}$. The point, $\mathbf{x}^*$, that maximizes the EI function is selected, the upper-scale model is evaluated using these parameter values, and the GP is retrained using the augmented training set. This process is repeated until the observed distance between the lower-scale model and upper-scale model is sufficiently small (i.e.\ $\Delta<\Delta_{th}$ where $\Delta_{th}$ is a threshold distance) or until a prescribed maximum number of iterations and the optimal set of parameters, denoted $X^+$ is obtained.

The approach is summarized in Algorithm \ref{Algorithm}.





{\fontsize{5}{5}\selectfont
\begin{algorithm}[!h]
\setstretch{1.2}
\nl: Run lower-scale model $\mathcal{L}$ to obtain reference solution $\mathcal{Z}^*_i$; \\
\nl: Generate $\mathcal{N}$ realizations of $X = \{\theta_1, \ldots,\theta_k; \xi_1, \dots, \xi_l\}$; \\

\nl: Run upper-scale model $\mathcal{M}$ for all $ \{X_j\}_{j=1}^{\mathcal{N}}$, to obtain $\mathcal{Z}_{i,j}$; \\

\nl: Project $ \{\mathcal{Z}_{i,j}\}_{j=1}^{\mathcal{N}}$ , $\mathcal{Z}^*_i$ onto $\mathcal{G}(\hat{p},n)$ and compute distances $d_i$ as in Eq.\ \eqref{eq:Grassmann_Distance} and any other distance measures of interest.

\nl: Compute the overall weighted distance measure, $\Delta_i, i=1,\dots,\mathcal{N}$ using Eq.\ \eqref{eqn:d_total} to form training dataset $\mathcal{D} = \{X_i,\Delta_i\}_{i=1}^{\mathcal{N}}$;

\nl: Step t=0; \ $\Delta_{th}$=$\epsilon$;

\nl: \textbf{while} ($\text{min} \{\Delta_i\}_{i=1}^{\mathcal{N}+t} >\Delta_{th}$ \& $t<T$) \textbf{do}

\nl: \ \ \ \ Construct GP surrogate as in Eq.\ \eqref{conditional} and generate $\mathcal{N}^{*}$ samples;

\nl: \ \ \ \ \textbf{for} all $i=1,\ldots,\mathcal{N}^{*} $ \textbf{do} \\

\nl: \ \ \ \ \ \ \ \ Compute distance $\Delta_i$ and $E[I(\mathbf{X}^{(i)})]$ as in Eq.\ \eqref{EI};

\nl: \ \ \ \ \textbf{end for}

\nl: \ \ \ \ Choose sample $X^*$ so that EI is maximized;

\nl: \ \ \ \ Solve upper-scale model $\mathcal{M}$ for $X^* = \{\theta_1, \ldots,\theta_k; \xi_1, \dots, \xi_l\}$ and compute $\Delta^*$;

\nl: \ \ \ \ Append $\{X^*,\Delta^*\}$ to dataset $\mathcal{D}$;

\nl: \ \ \ \ $t$ $\leftarrow$ $t+1$;

\nl: \textbf{end while}

\nl: Obtain optimal set of parameters $X^{+} = \text{argmin}\{\Delta(X)\} $;

    \caption{{Proposed Grassmannian EGO scheme} \label{Algorithm}}
\end{algorithm}
}

\section{Why perform manifold projection?}
Given that the central feature of the proposed method is a comparison between lower- and upper-scale model solutions in the form of matrix-valued snapshots, one might postulate that natural distance measures between such solutions can be obtained directly through standard norms. Why then perform a manifold projection, and in particular a Grassmannian projection? 

Indeed standard norms in the high-dimensional ambient space can be used to compare such field quantities. But, such comparisons may be less effective than the proposed manifold projection method for the following reasons:
\begin{figure}[ht!]
\begin{center}
\includegraphics[width=\textwidth]{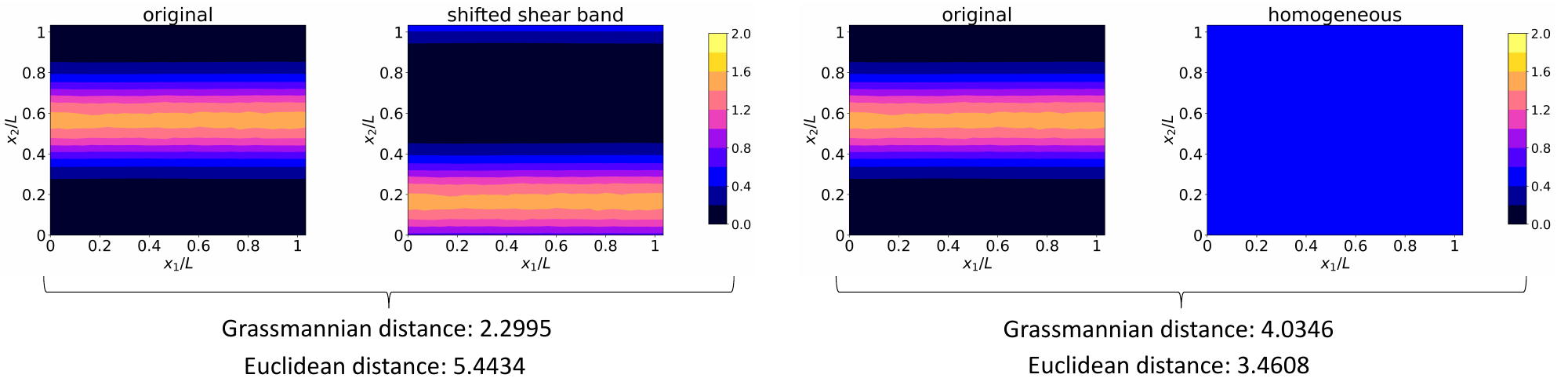}
\caption{Comparison of Grassmannian and Euclidean distances for different pairs of solutions. Left: the solutions both produce shear bands but in different locations. Right: one solution produces a shear band and the other homogeneous plastic strain. The Grassmannian distance identifies that the solutions with shear bands are closer, while the Euclidean distance suggests that the homogeneous solution is closer to the original than the solution with a shifted shear band. }
\label{fig:fcomp}
\end{center}
\end{figure}
\begin{itemize}
    \item It is known that Euclidean distance measures break down in high dimensions such that, as stated in \cite{aggarwal2001surprising} ``in high dimensional space, the concept of proximity, distance or nearest neighbor may not even be qualitatively meaningful.'' This can make identifying ``similar'' solutions a particular challenge since two solutions that have the similar structural features but differ in more subtle ways (e.g. a translation of a critical feature such as a shear band) may not be discernible using a Euclidean distance from two solutions with different structural features. This is illustrated in Figure \ref{fig:fcomp}. The left plot shows two solutions with shear bands in different locations. These solutions are structurally similar and represent the same type of material behavior. The right plot shows a solution with a shear band and a solution with constant (homogeneous) plastic strain. These solutions are structurally different. This structural difference is identified by the Grassmann distance, which is larger than the Grassmann distance between the two structurally similar solutions. But the Euclidean distance actually identifies the homogeneous strain field as being ``closer'' to the original field than the solution with a translated shear band. Qualitatively, we would expect two solutions with shear bands to be closer since they exhibit the same type of material behavior, even if at a different location.
    \item Data in the ambient space may have large degrees of redundancy. Certain features are strongly dependent on one another and are often manifestations of the same, or highly related, phenomena. Indeed, this arises due to the physical constraints imposed by the governing equations. Grassmannian projection reduces the data set to a small set of orthogonal “features” that can be separated and ordered by their prominence in the data. The proposed method compares these features directly.
    \item Distances in the ambient space may be sensitive to non-structural features of the data. For example, certain linear transformations (e.g. scaling) will dominate distance metrics in the ambient space. But these are not important for matching the “structural signatures” of the lower- and upper-scale models. This has been observed, for example, in the image processing literature where Grassmannian distances are used to classify images with different levels of illumination \cite{hamm2008grassmann}. If necessary, such factors can be accounted for with additional distance measures. For example, in the application that follows, we ensure proper scaling by additionally matching global stress-strain responses.
    \item Distances in the ambient space may be very sensitive to noise in the data. In our case, local fluctuations in the solution are far less important than the dominant features (i.e. the shear band in the subsequent application). When projecting on the Grassmannian, these local fluctuations are associated with the higher modes (with small singular values) and are therefore filtered out in the SVD truncation. Therefore, they do not influence the Grassmannian distance measure. Again, the ability to discriminate using the Grassmann distance from noisy data has been demonstrated in the image processing literature \cite{turaga2008statistical}.
\end{itemize}

\section{Application to Amorphous Solids} \label{Application}

\subsection{Components of the multi-scale model}

The proposed framework is applied to the study of inelastic deformation of a metallic glass (MG) alloy (a model amorphous solid). The constituents of the coarse-graining methodology (described in Section \ref{CG-intro}) for this materials system are as follows:
\begin{enumerate}
    \item \textit{Lower-scale Model:} Non-equilibrium molecular dynamics (NEMD) model of an MG under simple shear;
    \item \textit{Upper-scale Continuum Theory:} Shear transformation zone (STZ) theory of plasticity;
    \item \textit{Continuum numerical solver:} Eulerian finite difference solver for STZ dynamics;
    \item \textit{Hypothesized means of data transfer:} Affine transformation of averaged atomic potential energies to effective temperature
\end{enumerate}
These components are individually detailed in the following.


\subsubsection{Lower-scale model: Molecular dynamics simulation of a metallic glass} \label{MD_simulation}

\begin{figure}[ht!]
\begin{center}
\includegraphics[width=0.8\textwidth]{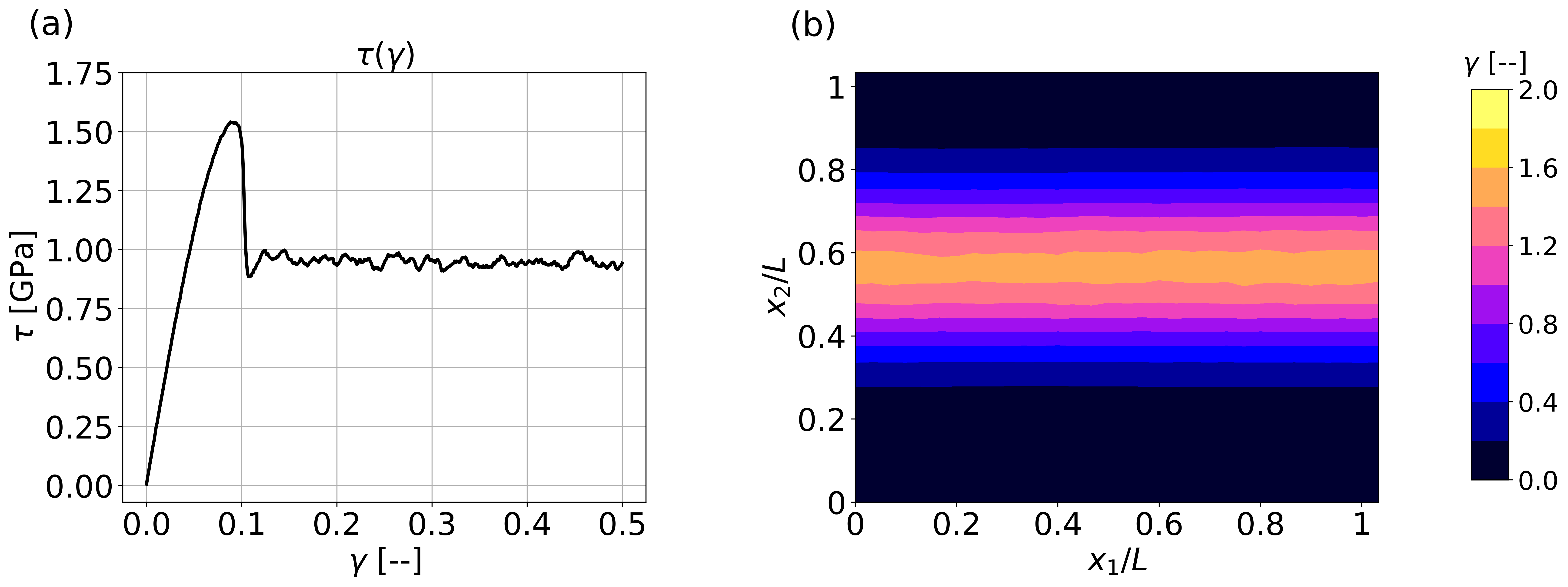}
\caption{Non-equilibrium molecular dynamics (NEMD) simulation of the Cu$_{50}$Zr$_{50}$ metallic glass system. Plot (a) shows the (shear) stress--strain curve and (b) the coarse-grained shear strain field at maximum net strain of $\gamma=50\%$ showing the formation of a thin shear band.\label{fig:MD-simulation}}
\end{center}
\end{figure}

The lower-scale model is a non-equilibrium molecular dynamics (NEMD) model executed with the LAMMPS software \cite{plimpton1995fast}, identical to the system studied by Hinkle et al.~\cite{hinkle2017coarse}. Atomic interactions are modeled using the well-known embedded-atom-method (EAM) interaction potential \cite{cheng2009atomic, sheng2011highly}. The glass was formed by taking a 50--50 composition of CuZr with 297,680 atoms and quenching the equilibrated liquid at a rate of \SI{1011}{\kelvin\per\second} to a temperature $T = \SI{100}{\kelvin}$. The material unit is a three dimensional rectangular prism with a thin third dimension ($\SI{400}{\angstrom} \times \SI{400}{\angstrom} \times \SI{30}{\angstrom}$) and fully periodic Lees--Edwards boundary conditions, allowing the system to be treated as effectively two-dimensional. Simple shear was applied to the quenched glass at constant volume and temperature, $T=\SI{100}{\kelvin}$, using the SLLOD equations of motion \cite{evans1984nonlinear} with a shear rate of $\dot{\gamma}=\SI{e-4}{\per\pico\second}$ and time step of \SI{0.005}{\pico\second}. The imposed shear induced the formation of a distinct shear band aligned with the direction of shear as illustrated in Figure \ref{fig:MD-simulation} (b). Such strain localization is a typical failure mode observed experimentally in metallic glasses and known to be reproducible in atomistic simulations \cite{shi2007evaluation,hufnagel2016deformation,alix2018shear}. In this simulation, the material exhibits classical stress--strain behavior for a MG under shear starting with linear elastic deformation, followed by yielding and a subsequent drop in the stress into a steady state stress plateau upon shear band formation and broadening, also shown in Figure \ref{fig:MD-simulation} (a).

\subsubsection{Upper-scale Continuum Theory: STZ Theory} \label{STZ}

A well-established theory to describe large-scale deformation in amorphous solids is the shear transformation zone (STZ) theory, which proposes that mechanically deformed amorphous solids dissipate plastic energy through the collective rearrangement of local clusters of atoms referred to as shear transformation zones, or STZs, \cite{bouchbinder2009nonequilibrium, falk2011deformation}. These STZs are believed to contain tens to hundreds of atoms, and are analogous in some ways to dislocation defects in crystalline solids \cite{manning2011vibrational, patinet2016connecting,cubuk2017structure,richard2020predicting}. The density of STZs in a glassy structure is quantified using an effective temperature \cite{langer2004dynamics, shi2007evaluation, bouchbinder2009nonequilibrium, falk2011deformation}. The concept of an effective temperature assumes that a non-equilibrium system in contact with a thermal reservoir can be decomposed into two weakly-coupled thermal subsystems: a fast kinetic/vibrational subsystem that equilibrates to the reservoir temperature on short timescales and a slow configurational subsystem that equilibrates on significantly longer timescales. The fast subsystem is characterized by the familiar thermal temperature $T$ whereas an effective temperature $T_\textit{eff}$ characterizes the slow subsystem. The effective temperature is defined as
\begin{equation}
    T_\textit{eff} = \frac{\partial U_C}{\partial S_C}
\end{equation}
where $U_C$ and $S_C$ are the energy and entropy of the slow configurational system, respectively. STZ models often use a dimensionless effective temperature defined as
\begin{equation}
    \chi \equiv \frac{k_B T_\textit{eff}}{E_Z}
\label{effective_temp}
\end{equation}
where $k_B$ is the Boltzmann factor and $E_Z$ is the energy required to create an STZ which is typically assumed to be $\sim\SI{2}{\electronvolt}$.


The STZ theory has undergone a number of changes since its introduction but fundamentally consists of two coupled equations -- one that describes the evolution of the plastic strain rate (given later for our system in Eqs.\ \eqref{dyn1}~\&~\eqref{dyn2}),
and an equation of motion for the effective temperature that describes the evolution of the structural/configurational state itself,
\begin{equation}
    c_0 \dot{\chi} = \widetilde{\mathbf{S}} : \mathbf{D}^{pl} (\chi_{\infty} - \chi) +  \nabla \cdot ( D_{\chi} \nabla \chi).
    \label{chi_eq1}
\end{equation}
The structural state of the system evolves according to a heat-like equation with a specific-heat-like parameter $c_0$ in the LHS of Eq.\ \eqref{chi_eq1}, which governs the amount of heat that flows into the slow thermal subsystem from mechanical work done on the system. The RHS of Eq.~\eqref{chi_eq1} consists of two terms. In the first term, $\widetilde{\mathbf{S}}$ denotes the de-dimensionalized deviatoric stress tensor, $\mathbf{D}^{pl}$ the rate of plastic deformation tensor and $\chi_{\infty}$ the steady-state effective temperature. This term indicates that the structural state of a driven glassy system evolves due to shear-induced plastic rearrangements, which drive the system towards a maximally disordered state. The second term corresponds physically to spatial diffusion of $\chi$, which works to relax effective temperature gradients. This term also includes the  effective temperature diffusivity $D_\chi$. In the section that follows, we introduce an STZ-based continuum solver and provide more details about the equations and the relevant parameters.

\subsubsection{Continuum numerical solver} \label{CM_model}


The continuum-scale STZ equations are solved with a fixed-grid, Eulerian finite difference numerical method for simulating quasi-static hypo-elastoplastic solids. The model has been used in prior studies to model simple shear deformation in Vitreloy and other MG systems \cite{rycroft2012fracture,vasoya16fracture}. Within the framework of hypo-elastoplasticity, the Eulerian rate-of-deformation tensor can be decomposed into an additive combination of elastic and plastic parts as $\mathbf{D} = \mathbf{D}^\textit{el} + \mathbf{D}^\textit{pl}$. This allows the linear-elastic constitutive relation to be rewritten as
\begin{equation}
    \frac{\mathcal{D}\mathbf{\boldsymbol\sigma}}{\mathcal{D}t} = \mathbf{C}:\mathbf{D}^\textit{el} = \mathbf{C}:(\mathbf{D}-\mathbf{D}^\textit{pl})
\label{constitutive}
\end{equation}
where $\mathbf{C}$ is the stiffness tensor for an isotropic and homogeneous medium and $\frac{\mathcal{D}\mathbf{\sigma}}{\mathcal{D}t}$ is the Truesdell objective stress rate. The velocity field evolves according to a continuum formulation of Newton's second law, $\rho\frac{d\mathbf{v}}{dt}= \nabla\cdot\boldsymbol{\sigma}$ where $\rho$ is the density of the material. In the limit of long times and correspondingly small deformation rates, this may be replaced by the constraint that the material maintains quasi-static equilibrium,
\begin{equation}
  \nabla \cdot \boldsymbol\sigma = \mathbf{0}.
\label{eqn:qs_constr}
\end{equation}
Since the MD simulations use a small domain size, and metallic glasses have large elastic wave speeds, it is valid to use the quasi-staticity assumption in this study.
The numerical implementation exploits an analogy between incompressible fluid dynamics and Eqs.~\eqref{constitutive}~\&~\eqref{eqn:qs_constr} developed by~\citet{rycroft2015eulerian}. In a first step, the $\mathbf{C} : \mathbf{D}$ term in Eq.~\eqref{constitutive} is neglected, and the resulting equation is advanced for a single timestep to compute an intermediate stress tensor. In a second step, the intermediate stress is orthogonally projected onto the manifold of solutions satisfying Eq.~\eqref{eqn:qs_constr} through the solution of an elliptic equation for the velocity.

An expression for the plastic part of the rate of deformation tensor is given by the STZ theory as
\begin{equation}
  \mathbf{D}^\textit{pl}=\begin{cases}
    \mathbf{0}, & \|\widetilde{\mathbf{S}}\|<1,\\
  \frac{\epsilon_0}{\tau_0}e^{-1/\chi}(1-\frac{1}{\|\widetilde{\mathbf{S}}\|}) \frac{\widetilde{\mathbf{S}}}{\|\widetilde{\mathbf{S}}\|} F, & \|\widetilde{\mathbf{S}}\| \ge 1,
  \end{cases}
\label{dyn1}
\end{equation}
where $F=F(\widetilde{\mathbf{S}})$ is a monotonic function of the deviatoric Cauchy stress  $\widetilde{\mathbf{S}}$, normalized in terms of the yield stress $s_y$. When $\|\widetilde{\mathbf{S}}\|<1$, the amorphous material behaves elastically, and $\mathbf{D}^\textit{pl}=\mathbf{0}$. In Eq.~\eqref{dyn1}, the parameter $\epsilon_0$ denotes the average number of atoms contained in a typical STZ, while $\tau_0$ is a molecular vibration timescale. Following Hinkle et al.~\cite{hinkle2017coarse}, the function $F$ is set to
\begin{equation}
    F = -2 + \|\widetilde{\mathbf{S}}\| + \exp(-\|\widetilde{\mathbf{S}}\|)(2+\|\widetilde{\mathbf{S}}\|).
\label{dyn2}
\end{equation}
As we have shown in the previous section, the effective temperature field evolves according to Eq.~\eqref{chi_eq1}. The effective temperature diffusivity is given by
\begin{equation}
    D_{\chi} = l_{\chi}^2\sqrt{(\mathbf{D}^\textit{pl}:\mathbf{D}^\textit{pl})}
\label{diffusivity}
\end{equation}
where $\l_{\chi}$ is the diffusion length scale. The dependence of $D_{\chi}$ on the rate of plastic deformation $\mathbf{D}^\textit{pl}$ shows that the diffusion rate of the material varies spatially and proportionally to the local rate of plastic deformation. Finally, we note that the local shear strain calculations are based on the Green-Lagrange strain tensor $\mathbf{E}$ which is computed separately, based on a procedure that can be found in \cite{rycroft2015eulerian}. Once the strain tensor is computed, the xy-component is used which essentially represents the strain caused by the simple shear.

A list of the continuum model parameters, with typical values, is provided in Table \ref{table:param}. 

Initialization of the model requires a spatial effective temperature field that defines the stress-free configurational structure of the system, which we will compute directly from an MD simulation. Typical continuum boundary conditions for the material velocity field correspond to clamped parallel plates, whereby a prescribed velocity is imposed on the top and bottom boundaries of the simulation domain. By contrast, the MD simulations in this work employ fully-periodic boundary conditions, and this discrepancy will affect the accuracy of the computed initial effective temperature field. To ensure consistency, we employ a two-dimensional formulation of the transformation methodology recently developed by~\citet{boffi20transform}. A fixed reference domain with coordinates $\mathbf{X}$ is mapped to the physical domain with coordinates $\mathbf{x}$ through a time-dependent transformation $\mathbf{T}(t)$, so that $\mathbf{x} = \mathbf{T}(t)\mathbf{X}$. The equations of motion are reformulated and solved directly in the $\mathbf{X}$ coordinates, and shear deformation is imposed through the transformation
\begin{equation}
  \mathbf{T}(t) = \begin{pmatrix} 1 & \lambda t \\ 0 & 1
\end{pmatrix}.
\end{equation}
This procedure decouples material deformation from material boundary conditions, and enables implementation of simple shear with periodic boundary conditions in our continuum simulation.

\begin{table}[ht!]
\small
\caption{Parameters of the STZ continuum model and corresponding typical values for a coarse-grained CuZr glass. The symbol ``\textendash" indicates dimensionless parameters.}
\vspace{-5pt}
\centering
\begin{tabular}{l c c c c}
\hline
\hline
PARAMETERS   & & UNIT &  Value \\ [0.5ex]
\hline
Elastic shear modulus    & \hspace{-30mm} $\mu$ &  GPa &    20 \vspace{-1pt} \\
Bulk modulus    & \hspace{-30mm} $K$ &  GPa &    66.23 \vspace{-1pt} \\
Molecular vibration timescale    & \hspace{-30mm} $\tau_0$ &  ps &    0.1 \vspace{-1pt} \\
Diffusion length scale    & \hspace{-30mm} $\ell_{\chi}$ &  \AA &    25.59 \vspace{-2pt} \\
Plastic work fraction    & \hspace{-30mm} $c_0$ &  \textendash &    0.3 \vspace{-1pt} \\
Typical size of an STZ    & \hspace{-30mm} $\epsilon_0$ &   \textendash &    44.53 \vspace{-1pt} \\
Yield stress  & \hspace{-30mm} $s_y$ & GPa & 0.85 \vspace{-1pt} \\
Steady-state effective temperature    & \hspace{-30mm} $\chi_{\infty}$ &   \textendash &    0.13 \vspace{-1pt} \\
STZ formation energy*    & \hspace{-30mm} $e_z/k_B$ &   K &    21,000 \vspace{-1pt} \\ [1ex]
\hline
\hline
{* Boltzmann constant $k_B=\SI{1.3806488d-23}{\joule \per \kelvin}$}
\end{tabular}
\label{table:param}
\end{table}

\subsubsection{Hypothesized coarse-graining protocol} \label{CG}

In this section, we describe the hypothesized operations ($\mathcal{C}$ and $\mathcal{T}$ from Section 3) that are used to coarse-grain data from MD simulations and pass this coarse-grained data to the continuum STZ model. The protocol, introduced by Hinkle et al. \cite{hinkle2017coarse}, first coarse-grains the atomic potential energies using a Gaussian moving average and then applies an affine relation, first  suggested by Shi et al.\ \cite{shi2007evaluation}, between average potential energy and $\chi$. We emphasize here that these relations are hypothesized and that the methodology presented here allows the straightforward consideration of alternate hypothesized relations and even the potential to learn these relations.

More specifically, the coarse-grained atomic potential energy at spatial location $\alpha$ is given by
\begin{equation}
    E_\alpha = \frac{\sum_n g_n E_n}{\sum_n g_n}
\label{coarse-grain}
\end{equation}
where $E_n$ is the atomic-energy of atom $n$, $g_n$ is the Gaussian windowing function
\begin{equation}
g_n = \frac{2}{\sqrt{2 \pi c}} \exp{\bigg( - \frac{{r_n}^2}{2 c^2} \bigg)}
\label{Gaussia_wind}
\end{equation}
that weights the contributions of each atom $n$ having a distance $r_n$ from grid point $\alpha$, and the summation is taken for all atoms in the neighborhood of $\alpha$ having $r_n \le r_\textit{cut}$. The length-scale parameter $c$ has been extensively studied by Hinkle et al.~\cite{hinkle2017coarse} with the primarily conclusion that there is a critical length-scale below which the coarse graining procedure breaks down. Here, we take $c=\SI{50}{\angstrom}$ as suggested by Hinkle et al.~\cite{hinkle2017coarse} although this parameter could be learned using the proposed methodology. Eq.\ \eqref{coarse-grain} represents the generic operation $\mathcal{C}$ presented earlier.

Next, the coarse-grained potential energy is mapped to effective temperature. Following Shi et al.~\cite{shi2007evaluation}, we assume that atomic potential energies can be linearly mapped onto effective temperatures as

\begin{equation}
    \chi_\alpha = \beta(E_\alpha-E_0)
\label{mapping}
\end{equation}
where $E_\alpha$ is the coarse-grained atomic potential energy from Eq.\ \eqref{coarse-grain}, $E_0$ is a reference potential energy per atom for an ideally-jammed glass, and $\beta$ is a scaling factor with units eV$^{-1}$. Eq.\ \eqref{mapping} represents the generic operation $\mathcal{T}$ presented earlier with transformation parameter vector $\Xi =\{\beta, E_0\}$.

\subsection{Problem Definition}

\subsubsection{Coarse-graining parameters}

The model presented above introduces a total of eight model parameters $\Theta=\{\mu, K, \epsilon_0, \tau_0, c_0, s_y, \chi_\infty, l_\chi\}$ and three coarse-graining parameters $\Xi=\{c, \beta, E_0\}$. The elastic modulus, $\mu$ and bulk modulus, $K$, can be directly inferred from the linear elastic response of the MD simulations and are set as in Table \ref{table:param}. A detailed sensitivity analysis (not presented) indicates that, in the regime of interest, the model is relatively insensitive to $\epsilon_0$ and  $\l_\chi$. Hence, these are assigned nominal values as given in Table \ref{table:param}. The value of the molecular vibration parameter $\tau_0$, is consistent with the one used in  \cite{hinkle2017coarse}. Finally, $\chi_\infty$ can be directly determined from prescribed values of coarse-graining parameters $\beta$ and $E_0$ and the coarse-graining length-scale $c=\SI{50}{\angstrom}$ is assigned from \cite{hinkle2017coarse} as discussed above. Thus, demonstration of the proposed coarse-graining methodology requires optimization of four parameters given by: $\{\Theta; \Xi\}=\{c_0, s_y; \beta, E_0\}$.

\subsubsection{Defining the objective function}
\label{sec:obj_func}

The four parameters for calibration are optimized using the Grassmannian EGO with an objective function defined to compare the MD simulations and the continuum model using the strain fields at several increments in the deformation and the global stress--strain response. The objective function is therefore a distance measure between MD and continuum response comprised of two components.

The first distance component, $d_{\bar{\Gamma}}$, accounts for the difference between the reference $\Gamma^{\textit{ref}}(\gamma)$ and observation $\Gamma(\gamma)$ strain fields at $n_{\gamma} = 9$ global strain values, $\gamma = \{0.02$, $0.04$, $0.06$, $0.08$, $0.09$, $0.1$, $0.2$, $0.35$, $0.5\}$. Values are chosen to capture model behavior in the linear-elastic regime, during stress-overshoot upon shear band formation, and in steady-state as shear bands broaden. Local strains at coarse-grained grid points from MD are computed using a weighted method based on a measure of nonaffine displacement modified from Falk and Langer \cite{falk1998dynamics} and detailed by Hinkle et al.~\cite{hinkle2017coarse}. The Grassmann distance between continuum and MD strain fields at each snapshot, $\{d_{{\mathcal{G}}_i}\}_{i=1}^{n_{\gamma}}$, are computed according to Eq.\ \eqref{eq:Grassmann_Distance}. The total distance between the strain fields $d_{\bar{\Gamma}}$, is taken as the average of distances from all nine snapshots,
\begin{equation}
  d_{\bar{\Gamma}} = \frac{1}{n_{\gamma}}\sum_{i=1}^{n_{\gamma}} d_{\mathcal{G}_i}.
  \label{distances1}
\end{equation}

As discussed in Section \ref{grass-proj}, the geodesic distance accounts for differences in the structural signature of the fields. The features that are captured by points on the Grassmannian (basis vectors) are among others the shape, location and relative magnitude of the shear strain. Moreover, discrepancies related to the absolute magnitude and angle of the shear band for example, are not encapsulated by the geodesic distance. 

The second component is based on the distance between the reference and observed stress--strain curves $(\tau-\gamma)$. Shear stress is calculated as the mean value $\{\bar{\tau}_i\}_{i=1}^{t_f}$, of the corresponding stress fields.
The distance is computed similarly as the average of the absolute differences at all $n_f=100$ simulation frames,
\begin{equation}
  d_{\bar{\tau}} = \frac{1}{n_f}\sum_{j=1}^{n_f} |\bar{\tau}^{\textit{ref}}_j - \bar{\tau}_{j}|.
  \label{distances2}
\end{equation}
We note that the introduction of this stress-strain measure, in addition to the Grassmannian measure, has the effect of acting as an ``external'' scaling measure (see Section 4). Although not acting directly on the scale of the local strains, it sets a scale for the overall stress and strain that serves to complement the feature recognition of the Grassmann distance. Finally, the distance function is computed as a combination of the strain field distance and the stress--strain distance as
\begin{equation}
    d_{\bar{\Gamma}+\bar{\tau}} = w_1 d_{\bar{\Gamma}} + w_2 d_{\bar{\tau}}.
\label{final_dist2}
\end{equation}
Sufficient regularization is achieved with the use of normalizing weights $(w_1,w_2)$, assigned to account for any scaling discrepancies.
Moreover, these weights regulate the desired degree of reproducibility for each of the various quantities of interest (QoIs).
For this application, we have chosen $(w_1,w_2)$ = $(0.25,0.75)$.

\subsection{Model validation} \label{Validation}




We first validate the proposed coarse-graining methodology by selecting as reference, the full-field solution obtained from a single forward continuum model evaluation having fixed (target) parameters as shown in Table \ref{table:validation}.

\begin{table}[ht!]
\small
\caption{Target values, distribution and optimized sample for the calibration parameters $\{\Theta; \Xi\}=\{c_0, s_y; \beta, E_0\}$.}
\vspace{-5pt}
\centering
\begin{tabular}{l c c c c c}
\hline
\hline
PARAMETERS & \hspace{43pt} & UNIT &  Target values & Distribution & Optimal sample, $X^{+}$ \\ [0.5ex]
\hline
Plastic work fraction    &  $c_0$ &  \textendash &    0.550  & $\sim \mathcal{U}(0.05, 1)$ & 0.612 \vspace{-1pt} \\
Yield stress  &  $s_y$ & GPa & 1.006 & $\sim \mathcal{U}(0.8, 1.15)$ & 1.002 \vspace{-1pt} \\
Scaling factor    &  $\beta$ &   eV$^{-1}$ &8.277 & $\sim \mathcal{U}(2, 15)$ & 8.710 \vspace{-1pt} \\
Reference potential energy   &  $E_0$ &  eV &    -3.368 & $\sim \mathcal{U}(-3.390, -3.355)$ & -3.368 \vspace{-1pt} \\ [1ex]
\hline
\hline
\multicolumn{6}{l}{\footnotesize *$\mathcal{U}\left(a,b\right)$ denotes a uniform distribution with lower bound $a$ and upper bound $b$.}
\end{tabular}
\label{table:validation}
\end{table}

\begin{figure}[!ht]
\centering
\begin{minipage}{.4\textwidth}
  \centering
  \includegraphics[width=\textwidth]{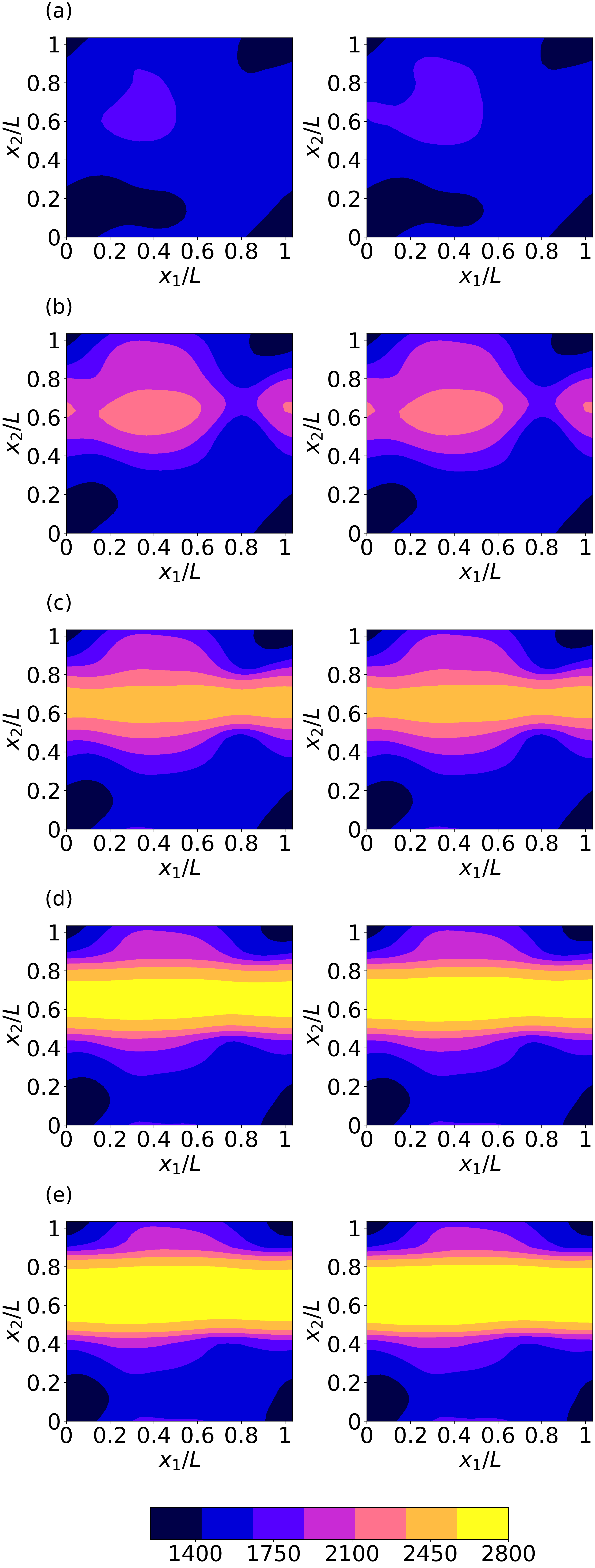}
  \captionof{figure}{Effective temperature fields of the reference simulation (left column) $T^\textit{ref}(\gamma)$ and the optimized continuum model simulation (right column) $T^\textit{CM}(\gamma)$ at shear strain values: (a) $9\%$, (b) $15\%$, (c) $20\%$, (d) $35\%$ and (e) $50\%$.}
  \label{fig:eff-temp-val}
\end{minipage}%
\hspace{20pt}
\begin{minipage}{.4\textwidth}
  \centering
  \includegraphics[width=\textwidth]{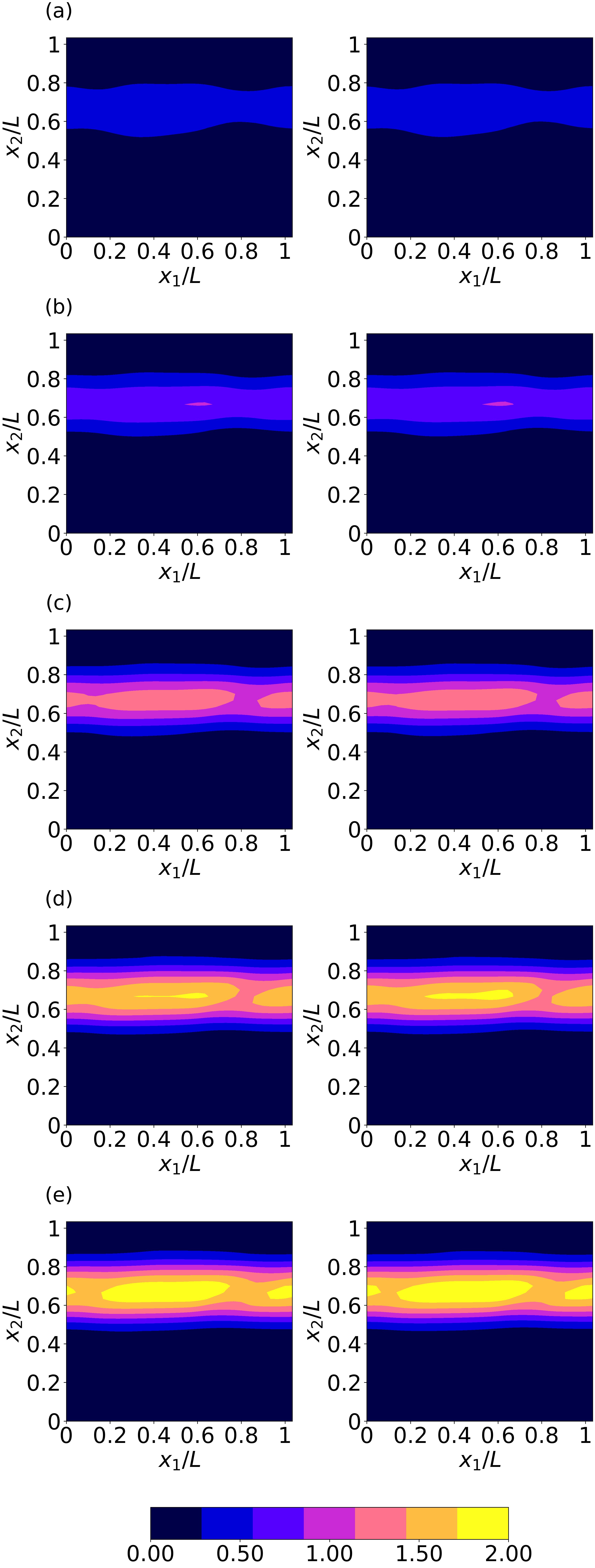}
  \captionof{figure}{Shear strain fields of the reference simulation (left column) $\Gamma^\textit{ref}(\gamma)$ and the optimized continuum model simulation (right column) $\Gamma^\textit{CM}(\gamma)$ at shear strain values: (a) $18\%$, (b) $25\%$, (c) $35\%$, (d) $45\%$ and (e) $50\%$.}
  \label{fig:shear-strain-val}
\end{minipage}
\end{figure}

\begin{figure}[!ht]
\begin{center}
\includegraphics[width=0.4\textwidth]{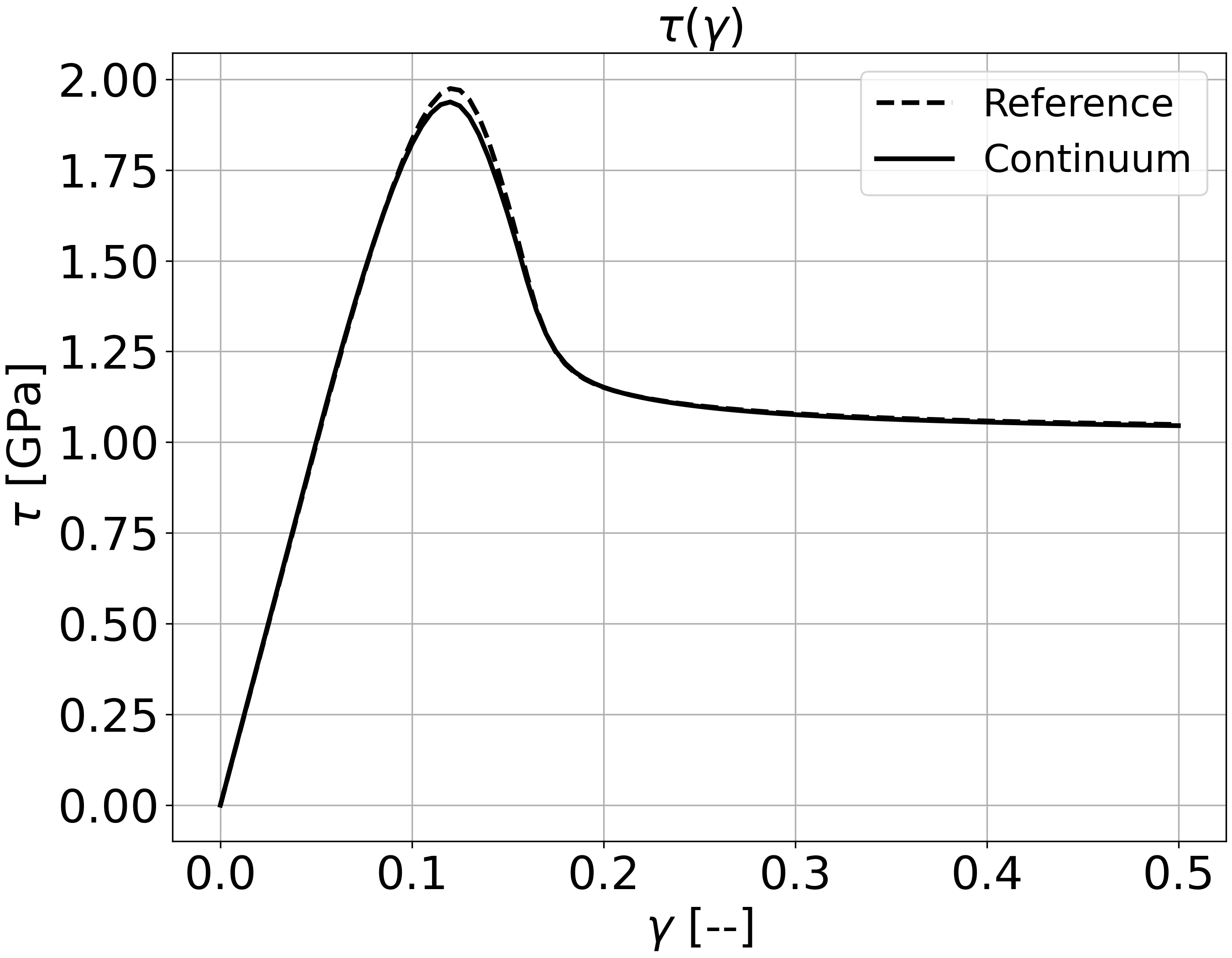}
\caption{Comparison of the shear stress--strain curve of the reference and optimized continuum STZ model simulation corresponding to an optimal set of input parameters $X^{+}=\{c_0, s_y; \beta, E_0\} = \{0.612, 1.002; 8.710, -3.368\}$ (see Table \ref{table:validation} for physical units).}
\label{fig:stress-val}
\end{center}
\end{figure}

A 2D simulation cell of an amorphous solid is considered, having dimensions of $\SI{400}{\angstrom} \times \SI{400}{\angstrom}$ and a grid resolution of $32 \times 32$. The analysis is performed in $n_f = 100$ simulation frames. Fully-periodic boundary conditions are imposed with constant velocity $\mathbf{v}(\mathbf{x},t)$ imposed at the top and bottom of the cell in the direction of shear having a shear-strain rate $\dot{\gamma}=\SI{e8}{\per\second}$, a value which is scaled appropriately to match the simulation units of the numerical solver. Target parameters have been chosen such that, at maximum shear strain $\gamma_{\max} = 50 \%$, a thin shear band forms approximately at the middle of the simulation box. Field quantities of interest are the effective temperature field $T(\gamma)$ and the strain field $\Gamma(\gamma)$ shown in the left columns of Figures \ref{fig:eff-temp-val} and \ref{fig:shear-strain-val}, although we only explicitly match the strain field in the learning. We also aim to match the shear stress--strain curve $(\tau-\gamma)$ shown in Figure \ref{fig:stress-val}(a).

An initial $\mathcal{N}=400$ training data are sampled from the parameter vector $X=\{\Theta; \Xi\}=\{c_0, s_y; \beta, E_0\}$ using Latin Hypercube Sampling (LHS). Parameters are treated as independent uniformly distributed random variables with bounds given in Table \ref{table:validation}. For each parameter sample, the continuum model is evaluated and the distance function is evaluated as described in Section \ref{sec:obj_func}. Computed distances $\{d_{\bar{\Gamma}+\bar{\tau}}\}_{i=1}^{\mathcal{N}}$, serve as labels $\Delta_i$, of the corresponding $X_i$ observations and GP model is trained for the distance function.

\begin{figure}[!ht]
\begin{center}
\includegraphics[width=0.9\textwidth]{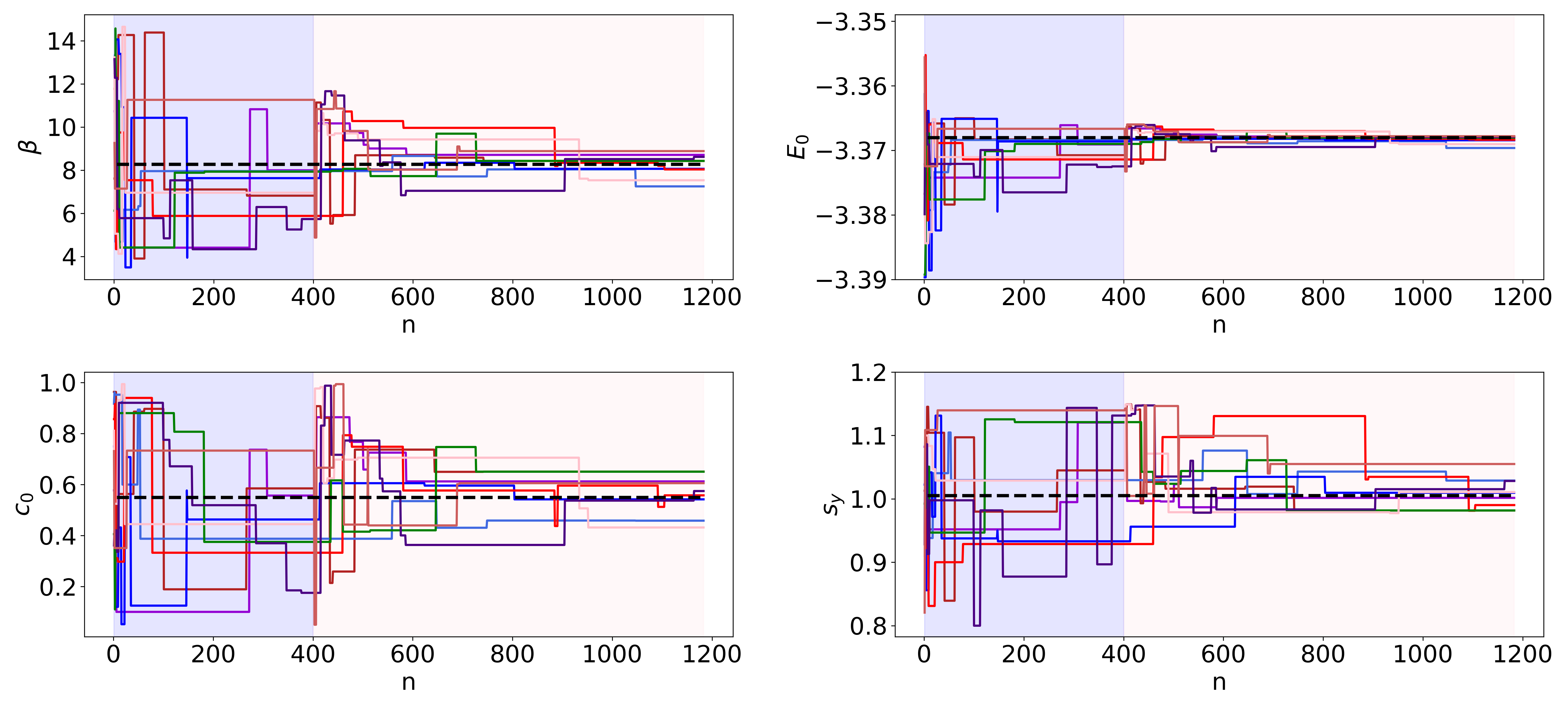}
\caption{Convergence of the four calibration parameters $X=\{\Theta; \Xi\}=\{c_0, s_y; \beta, E_0\}$ to the target values (dashed line) (see Table \ref{table:validation}) for 9 implementations of Grassmannian EGO each with a different random seed number. A total of 800 EGO iterations have resulted in a good convergence of the model parameters to the ``true'' values. }
\label{fig:convergence-val}
\end{center}
\end{figure}

A total of 800 EGO iterations are performed such that each subsequent sample is selected as the one that maximizes the EI function, yielding a total of 1200 inexpensive continuum model simulations and an estimate of the parameters $X^+$ (Table \ref{table:validation}) that provide the best match. This process is repeated 9 times using different random seed values to ensure reproducibility. Parameter convergence is plotted in Figure \ref{fig:convergence-val} for all 9 repeats showing a consistent convergence to the true parameter values of the reference. As explained, a sensitivity analysis is considered important, to quantify the model's sensitivity to parameters. We have found that the STZ model is more sensitive to some parameters (e.g., $E_0$) than others (e.g., $c_0$). As a result the convergence rate of optimal to reference values varies across parameters. However, it is expected that as additional samples are added for the construction of the surrogate model, the algorithm will eventually converge to the `best' values. Moreover, the right hand column of plots in Figures \ref{fig:eff-temp-val} and \ref{fig:shear-strain-val} show that the effective temperature and strain fields from the optimized model match those of the reference model nearly perfectly. Finally, Figure \ref{fig:stress-val} shows that the stress--strain curves for the optimized model and the reference model match almost perfectly.

\subsection{Results} \label{Results}

We now present the results of the Grassmannian EGO applied to the coarse-grained STZ model with reference given by the NEMD simulation of the CuZr glass discussed in Section \ref{MD_simulation}. In contrast to the model validation case, target parameter values are not available and thus the optimal parameters are evaluated based on the ability of Grassmannian EGO to reproduce the coarse-grained MD results.

\begin{figure}[ht!]
\centering
\begin{minipage}{.4\textwidth}
  \centering
  \includegraphics[width=\textwidth]{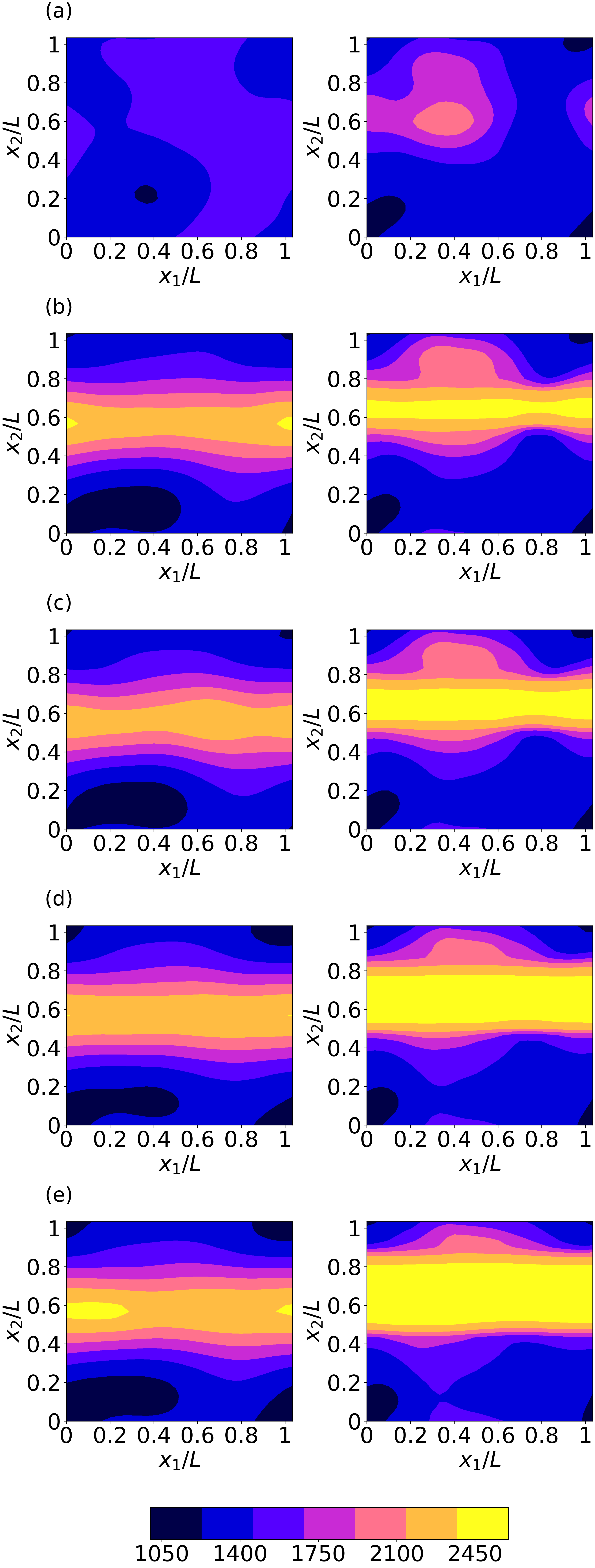}
  \captionof{figure}{Effective temperature fields of the reference MD simulation (left column) $T^\textit{ref}(\gamma)$ and the optimized continuum model simulation (right column) $T^\textit{CM}(\gamma)$ at shear strain values: (a) $9\%$, (b) $15\%$, (c) $20\%$, (d) $35\%$ and (e) $50\%$.}
  \label{fig:eff-temp-appl}
\end{minipage}%
\hspace{20pt}
\begin{minipage}{.4\textwidth}
  \centering
  \includegraphics[width=\textwidth]{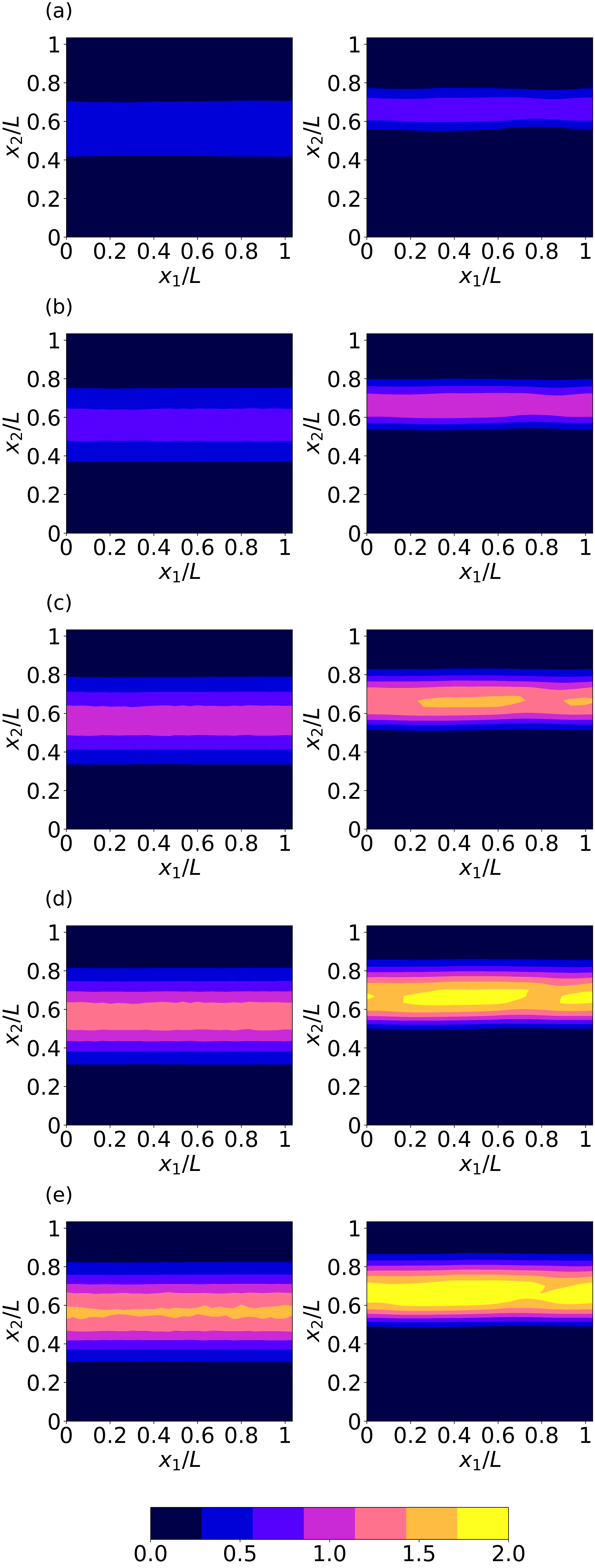}
  \captionof{figure}{Shear strain fields of the reference MD simulation (left column) $\Gamma^\textit{ref}(\gamma)$ and the optimized continuum model simulation (right column) $\Gamma^\textit{CM}(\gamma)$ at shear strain values: (a) $18\%$, (b) $25\%$, (c) $35\%$, (d) $45\%$ and (e) $50\%$.}
  \label{fig:shear-strain-appl}
\end{minipage}
\end{figure}

We feed the algorithm with $\mathcal{N}=350$ training observations generated with LHS. The bounds of the corresponding uniform distributions from which samples are drawn, are the same as in the validation case (Table \ref{table:validation}). Supervised learning is employed on the training dataset of computed distances $\{d_{\bar{\Gamma}+\bar{\tau}}\}_{i=1}^{\mathcal{N}}$ which serve as labels $\Delta_i$ and the corresponding $X_i$ observations from which the GP surrogate is trained. After the completion of 1835 EGO iterations, the algorithm identifies an optimal sample $X^{+}=\{c_0, s_y; \beta, E_0\} = \{0.164, 0.906; 6.882, -3.369\}$ after which there is no improvement.

In Figure \ref{fig:eff-temp-appl}, the effective temperature fields for both the reference coarse-grained MD and the optimal continuum are presented for various strain levels. Although we do not explicitly aim to match the effective temperature fields by incorporating their similarity into the distance metric, we find that the response between the models is similar at the different stages of deformation. We observe that the optimal solution of the continuum model produces a shear-band of higher magnitude which tends to increase in width as the simulation reaches maximum net strain (i.e.\ $\gamma=50\%$). The location and magnitude of the shear band highly depends on the initial effective temperature profile, which in turn depends on the coarse-grained atomic potential energy field and its linear mapping via the transformation parameters $\{\Xi\}=\{ \beta, E_0\}$ (Eq.\ \eqref{mapping}). We hypothesize that perhaps a non-linear mapping could yield to a more appropriate initial temperature field to be mapped to the continuum model that could potentially produce temperature fields of greater parity. Regardless, the match between the evolved effective temperature fields is deemed to be quite good and represents a notable improvement over previous calibration efforts. Discovery of an improved (non-linear) coarse-graining transformation remains an open research question.

\begin{figure}[!ht]
\begin{center}
\includegraphics[width=0.4\textwidth]{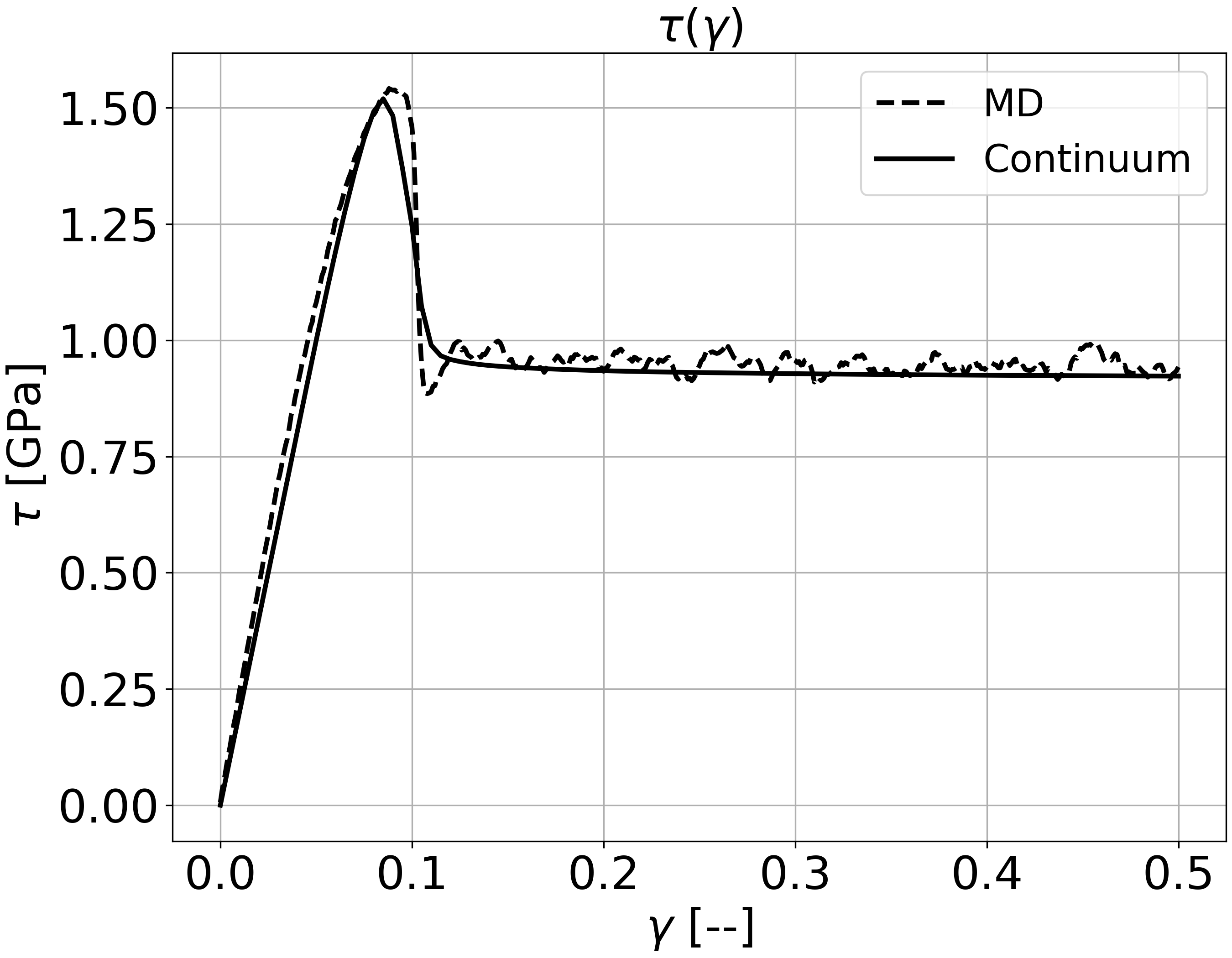}
\caption{Comparison of the stress--strain curve between the reference MD simulation of the Cu$_{50}$Zr$_{50}$ metallic glass and the continuum STZ model response for the optimal set of parameters $X^{+}=\{c_0, s_y; \beta, E_0\} = \{0.164, 0.906; 6.882, -3.369\}$ (see Table \ref{table:validation} for physical units).}
\label{fig:stress-vs-strain-appl}
\end{center}
\end{figure}

Comparison between the shear-strain fields of the MD and optimized continuum model is presented in Figure \ref{fig:shear-strain-appl}. We observe that the models exhibit very similar characteristic features, a prominent centralized shear band surrounded by regions of smaller homogeneous strains. As discussed, the Grassmannian distance between the two fields accounts for differences in the shape, location and relative magnitude of the strain fields. The continuum model clearly favors the formation of thinner shear bands with larger, more concentrated shear localization. From an early stage of the simulation the MD response is more diffuse with a gradual transition between regions of jammed and flowing material, whereas the optimal continuum exhibits a more abrupt transition. Discrepancies between the strain fields may be caused by a combination of factors including the imperfect coarse-graining hypothesis, general model-form error in the STZ theory (i.e.\ STZ theory does not perfectly model the complex dynamics that is captured in detailed MD), and the dual objective defined in the calibration (i.e.\ the trade-off between trying to to simultaneously match the strain field evolution and stress--strain curve). Nonetheless, the strain field matches quite well, and again represents significant improvement over previous efforts at calibrating the STZ model from MD data.

Comparison of the MD and calibrated STZ model stress--strain curves in Figure \ref{fig:stress-vs-strain-appl} shows a very close match. The STZ model is able to capture the linear elastic regime, stress overshoot and steady-state of the MD simulation very well.

\begin{figure}[!ht]
\begin{center}
\includegraphics[width=0.9\textwidth]{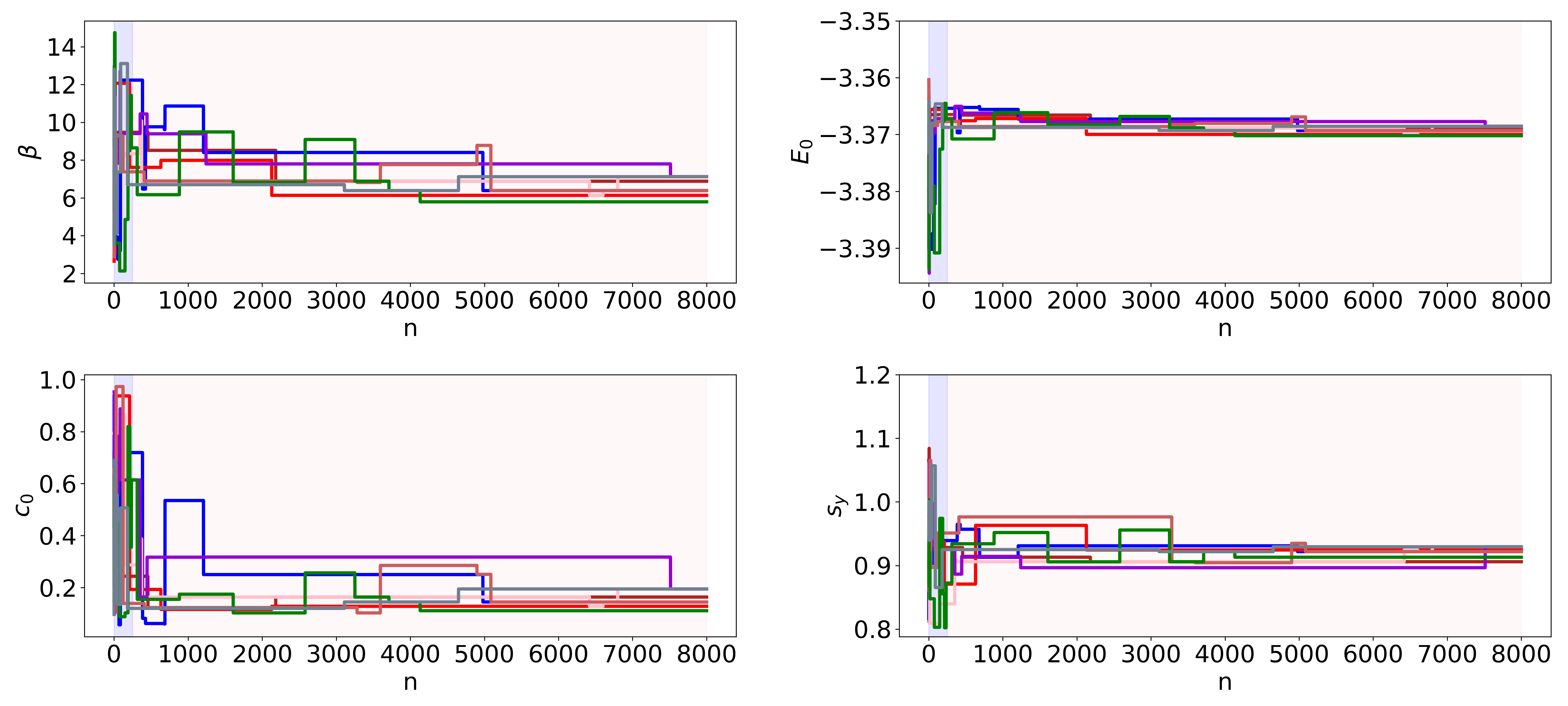}
\caption{Convergence of the four calibration parameters $X=\{\Theta; \Xi\}=\{c_0, s_y; \beta, E_0\}$ for 8 implementations of Grassmannian EGO each with a different random seed number. A total of 8,000 EGO iterations have resulted in the convergence of the parameters to a set of values with small uncertainty.}
\label{fig:convergence-appl}
\end{center}
\end{figure}

Similar to the previous section, the process is repeated 8 times to ensure reproducibility of results. We found that a large number of combinations of input parameters are capable of yielding a similar model response (equivalently similar manifold distance) and thus Grassmannian EGO occasionally (i.e. for certain random seed numbers) requires a large number of iterations in order to converge to the optimal set of parameters. We therefore run all 8 repeated calibrations for a total of 8,000 EGO iterations. In Figure \ref{fig:convergence-appl}, we observe convergence to a unique set of values with very small uncertainty. Grassmannian EGO was able to successfully parameterize the continuum model whose optimal response achieved a high level of similarity with the atomistic simulation.

Lastly we emphasize that, although the results shown herein do not show perfect agreement between atomistic and continuum models, these discrepancies are not a result of inadequate calibration by the Grassmannian EGO. Rather, these discrepancies results from intrinsic differences between the MD and continuum models. The proposal algorithm clearly converges to the best possible fit between these models. But, in the end, these models cannot produce a perfect match. To achieve a better fit, either the STZ model or the hypothesized coarse-graining methodology need to be improved. Improvements in both are topics of active research. Moreover, differences in the solution resulting from small variations in the optimized parameters have very small effect on the Grassmannian distance. Perhaps more telling, the strain field solutions from all eight optimization cases are visually nearly indistinguishable, pointing to the ability of the Grassmannian metric to compare structural features that are visually observable.

\section{Discussion and Conclusions}

In this work we introduce a generalized manifold learning framework, called Grassmannian EGO, for probabilistic learning of continuum models, (e.g.\ non-linear PDEs) based on atomistic simulation (e.g.\ molecular dynamics) data. A training dataset, formed by a carefully selected number of upper-scale model parameters and a distance metric that incorporates discrepancies between QoIs generated by the multiscale models, is used to train a Gaussian process surrogate. Projection of the QoIs in the form of high-dimensional data onto a Riemannian manifold allows us to construct their latent denoised representation, which is then used to feed the surrogate modeling algorithm with meaningful distances (i.e.\ labels). The objective of matching the response of the multiscale models is achieved by performing optimization on the manifold until an optimal sample is identified whose distance from a reference response is minimized. We demonstrate the ability of this methodology to transfer information from lower to higher scales and to achieve a high level of parity between the multiscale models. We ensure reproducibility of results by repeating the process each time by feeding the algorithm with different sets of training data.

Grassmannian EGO is computationally fast as it requires a small number of forward model evaluations. Even though the process of constructing a Gaussian process surrogate can be expensive as its covariance function scales quadratically with the number of added samples, local-surrogate updating algorithms can be employed to reduce the computational burden. Such algorithms can be used to construct local GP models, trained on small sets of nearest neighboring samples, since a new added sample generated by EGO influences the metamodel's prediction only at a certain region around it and not at the entire parametric space. Although not presented in this paper for the sake of brevity, a similar algorithm was used which resulted in a significant reduction of the computational cost.

The proposed methodology is first validated and then applied for the optimization of the parameters of a continuum model based on the shear transformation zone (STZ) theory of plasticity to reproduce mechanical deformation shown in a molecular dynamics simulation of a quenched CuZr metallic glass system. As demonstrated, the method is able to successfully calibrate the STZ model to reproduce the behavior of the metallic glass response from the MD simulation. Although the assumption of a linear mapping of the potential energy to the effective temperature was adopted in this study, non-linear mappings could be employed in a straightforward manner and different hypothesized course-graining transformations can be considered. The methodology can be extended and employed to verify consistency of the parameter calibration in the case where multiple MD simulations are generated from the same material composition. Once verified, the method can be employed for coarse-graining atomistic simulations of various material compositions, number of atoms and geometries. Furthermore, it can be extended and applied to three-dimensional atomistic simulation data, in which case, the manifold learning component can be treated with modern tensor factorization techniques  \cite{vesselinov2019unsupervised}.

The proposed scale-bridging approach is robust and capable of replacing prohibitively expensive atomistic simulations with computationally fast continuum model ones. This work is an early step towards using probabilistic manifold learning for coarse-graining lower-scale simulations and even though in this paper it was applied for the study of amorphous solids, we anticipate that it can find use in studies related to coarse-graining for different classes of materials (e.g.\ crystalline, non-crystalline, polymers, bio-materials and other soft materials.).

\section{Acknowledgements}
This project has been supported by the Department of Energy under project number DE-SC0020428. The project has also received support from the National Science Foundation under Grants No.~DMR-1910066 and No.~DMR-1909733. C.H.R.~was partially supported by the Applied Mathematics Program of the US DOE Office of Science Advanced Scientific Computing Research
under Contract No. DE-AC02-05CH11231. Molecular dynamics calculations were conducted using computational resources at the Maryland Advanced Research Computing Center (MARCC).

\bibliography{sample}

\begin{thebibliography}{78}
\providecommand{\natexlab}[1]{#1}
\providecommand{\url}[1]{\texttt{#1}}
\expandafter\ifx\csname urlstyle\endcsname\relax
  \providecommand{\doi}[1]{doi: #1}\else
  \providecommand{\doi}{doi: \begingroup \urlstyle{rm}\Url}\fi

\bibitem[Miller et~al.(1998)Miller, Tadmor, Phillips, and
  Ortiz]{miller1998quasicontinuum}
R~Miller, EB~Tadmor, R~Phillips, and M~Ortiz.
\newblock Quasicontinuum simulation of fracture at the atomic scale.
\newblock \emph{Modelling and Simulation in Materials Science and Engineering},
  6\penalty0 (5):\penalty0 607, 1998.

\bibitem[Noid(2013)]{noid2013perspective}
William~George Noid.
\newblock Perspective: Coarse-grained models for biomolecular systems.
\newblock \emph{The Journal of Chemical Physics}, 139\penalty0 (9):\penalty0
  09B201\_1, 2013.

\bibitem[Akkermans and Briels(2001)]{akkermans2001structure}
Reinier~LC Akkermans and Willem~J Briels.
\newblock A structure-based coarse-grained model for polymer melts.
\newblock \emph{The Journal of Chemical Physics}, 114\penalty0 (2):\penalty0
  1020--1031, 2001.

\bibitem[Li and Takada(2010)]{li2010characterizing}
Wenfei Li and Shoji Takada.
\newblock Characterizing protein energy landscape by self-learning multiscale
  simulations: Application to a designed $\beta$-hairpin.
\newblock \emph{Biophysical Journal}, 99\penalty0 (9):\penalty0 3029--3037,
  2010.

\bibitem[Toth(2007)]{toth2007effective}
Gergely Toth.
\newblock Effective potentials from complex simulations: a potential-matching
  algorithm and remarks on coarse-grained potentials.
\newblock \emph{Journal of Physics: Condensed Matter}, 19\penalty0
  (33):\penalty0 335222, 2007.

\bibitem[Reith et~al.(2003)Reith, P{\"u}tz, and
  M{\"u}ller-Plathe]{reith2003deriving}
Dirk Reith, Mathias P{\"u}tz, and Florian M{\"u}ller-Plathe.
\newblock Deriving effective mesoscale potentials from atomistic simulations.
\newblock \emph{Journal of Computational Chemistry}, 24\penalty0 (13):\penalty0
  1624--1636, 2003.

\bibitem[Lyubartsev and Laaksonen(1995)]{lyubartsev1995calculation}
Alexander~P Lyubartsev and Aatto Laaksonen.
\newblock Calculation of effective interaction potentials from radial
  distribution functions: A reverse {Monte} {Carlo} approach.
\newblock \emph{Physical Review E}, 52\penalty0 (4):\penalty0 3730, 1995.

\bibitem[Li et~al.(2016)Li, Bian, Yang, and Karniadakis]{li2016comparative}
Zhen Li, Xin Bian, Xiu Yang, and George~Em Karniadakis.
\newblock A comparative study of coarse-graining methods for polymeric fluids:
  {Mori--Zwanzig} vs. iterative {Boltzmann} inversion vs. stochastic parametric
  optimization.
\newblock \emph{The Journal of Chemical Physics}, 145\penalty0 (4):\penalty0
  044102, 2016.

\bibitem[Brini and Van~der Vegt(2012)]{brini2012chemically}
E~Brini and NFA Van~der Vegt.
\newblock Chemically transferable coarse-grained potentials from conditional
  reversible work calculations.
\newblock \emph{The Journal of Chemical Physics}, 137\penalty0 (15):\penalty0
  154113, 2012.

\bibitem[Savelyev and Papoian(2009)]{savelyev2009molecular}
Alexey Savelyev and Garegin~A Papoian.
\newblock Molecular renormalization group coarse-graining of polymer chains:
  application to double-stranded {DNA}.
\newblock \emph{Biophysical Journal}, 96\penalty0 (10):\penalty0 4044--4052,
  2009.

\bibitem[Chaimovich and Shell(2011)]{chaimovich2011coarse}
Aviel Chaimovich and M~Scott Shell.
\newblock Coarse-graining errors and numerical optimization using a relative
  entropy framework.
\newblock \emph{The Journal of Chemical Physics}, 134\penalty0 (9):\penalty0
  094112, 2011.

\bibitem[Carmichael and Shell(2012)]{carmichael2012new}
Scott~P Carmichael and M~Scott Shell.
\newblock A new multiscale algorithm and its application to coarse-grained
  peptide models for self-assembly.
\newblock \emph{The Journal of Physical Chemistry B}, 116\penalty0
  (29):\penalty0 8383--8393, 2012.

\bibitem[Izvekov and Voth(2005)]{izvekov2005multiscale}
Sergei Izvekov and Gregory~A Voth.
\newblock A multiscale coarse-graining method for biomolecular systems.
\newblock \emph{The Journal of Physical Chemistry B}, 109\penalty0
  (7):\penalty0 2469--2473, 2005.

\bibitem[Izvekov et~al.(2005)Izvekov, Violi, and Voth]{izvekov2005systematic}
Sergei Izvekov, Angela Violi, and Gregory~A Voth.
\newblock Systematic coarse-graining of nanoparticle interactions in molecular
  dynamics simulation.
\newblock \emph{The Journal of Physical Chemistry B}, 109\penalty0
  (36):\penalty0 17019--17024, 2005.

\bibitem[Villet and Fredrickson(2010)]{villet2010numerical}
Michael~C Villet and Glenn~H Fredrickson.
\newblock Numerical coarse-graining of fluid field theories.
\newblock \emph{The Journal of Chemical Physics}, 132\penalty0 (3):\penalty0
  034109, 2010.

\bibitem[Marrink et~al.(2007)Marrink, Risselada, Yefimov, Tieleman, and
  De~Vries]{marrink2007martini}
Siewert~J Marrink, H~Jelger Risselada, Serge Yefimov, D~Peter Tieleman, and
  Alex~H De~Vries.
\newblock The martini force field: coarse grained model for biomolecular
  simulations.
\newblock \emph{The Journal of Physical Chemistry B}, 111\penalty0
  (27):\penalty0 7812--7824, 2007.

\bibitem[Gooneie et~al.(2017)Gooneie, Schuschnigg, and
  Holzer]{gooneie2017review}
Ali Gooneie, Stephan Schuschnigg, and Clemens Holzer.
\newblock A review of multiscale computational methods in polymeric materials.
\newblock \emph{Polymers}, 9\penalty0 (1):\penalty0 16, 2017.

\bibitem[Ruiz et~al.(2015)Ruiz, Xia, Meng, and Keten]{ruiz2015coarse}
Luis Ruiz, Wenjie Xia, Zhaoxu Meng, and Sinan Keten.
\newblock A coarse-grained model for the mechanical behavior of multi-layer
  graphene.
\newblock \emph{Carbon}, 82:\penalty0 103--115, 2015.

\bibitem[Homer and Schuh(2009)]{homer2009mesoscale}
Eric~R Homer and Christopher~A Schuh.
\newblock Mesoscale modeling of amorphous metals by shear transformation zone
  dynamics.
\newblock \emph{Acta Materialia}, 57\penalty0 (9):\penalty0 2823--2833, 2009.

\bibitem[Hinkle et~al.(2017)Hinkle, Rycroft, Shields, and
  Falk]{hinkle2017coarse}
Adam~R Hinkle, Chris~H Rycroft, Michael~D Shields, and Michael~L Falk.
\newblock Coarse graining atomistic simulations of plastically deforming
  amorphous solids.
\newblock \emph{Physical Review E}, 95\penalty0 (5):\penalty0 053001, 2017.

\bibitem[Chmiela et~al.(2017)Chmiela, Tkatchenko, Sauceda, Poltavsky,
  Sch{\"u}tt, and M{\"u}ller]{chmiela2017machine}
Stefan Chmiela, Alexandre Tkatchenko, Huziel~E Sauceda, Igor Poltavsky,
  Kristof~T Sch{\"u}tt, and Klaus-Robert M{\"u}ller.
\newblock Machine learning of accurate energy-conserving molecular force
  fields.
\newblock \emph{Science Advances}, 3\penalty0 (5):\penalty0 e1603015, 2017.

\bibitem[Han et~al.(2017)Han, Zhang, Car, et~al.]{han2017deep}
Jiequn Han, Linfeng Zhang, Roberto Car, et~al.
\newblock Deep potential: A general representation of a many-body potential
  energy surface.
\newblock \emph{arXiv preprint arXiv:1707.01478}, 2017.

\bibitem[Zhang et~al.(2018{\natexlab{a}})Zhang, Han, Wang, Car, and
  E]{zhang2018deep}
Linfeng Zhang, Jiequn Han, Han Wang, Roberto Car, and Weinan E.
\newblock Deep potential molecular dynamics: a scalable model with the accuracy
  of quantum mechanics.
\newblock \emph{Physical Review Letters}, 120\penalty0 (14):\penalty0 143001,
  2018{\natexlab{a}}.

\bibitem[Zhang et~al.(2018{\natexlab{b}})Zhang, Han, Wang, Car, and
  E]{zhang2018deepcg}
Linfeng Zhang, Jiequn Han, Han Wang, Roberto Car, and Weinan E.
\newblock {DeePCG}: Constructing coarse-grained models via deep neural
  networks.
\newblock \emph{The Journal of Chemical Physics}, 149\penalty0 (3):\penalty0
  034101, 2018{\natexlab{b}}.

\bibitem[Moradzadeh and Aluru(2019)]{moradzadeh2019transfer}
Alireza Moradzadeh and Narayana~R Aluru.
\newblock Transfer-learning-based coarse-graining method for simple fluids:
  Toward deep inverse liquid-state theory.
\newblock \emph{The Journal of Physical Chemistry Letters}, 10\penalty0
  (6):\penalty0 1242--1250, 2019.

\bibitem[Goodfellow et~al.(2014)Goodfellow, Pouget-Abadie, Mirza, Xu,
  Warde-Farley, Ozair, Courville, and Bengio]{goodfellow2014generative}
Ian Goodfellow, Jean Pouget-Abadie, Mehdi Mirza, Bing Xu, David Warde-Farley,
  Sherjil Ozair, Aaron Courville, and Yoshua Bengio.
\newblock Generative adversarial nets.
\newblock In \emph{Advances in Neural Information Processing Systems}, pages
  2672--2680, 2014.

\bibitem[Durumeric and Voth(2019)]{durumeric2019adversarial}
Aleksander~EP Durumeric and Gregory~A Voth.
\newblock Adversarial-residual-coarse-graining: Applying machine learning
  theory to systematic molecular coarse-graining.
\newblock \emph{The Journal of Chemical Physics}, 151\penalty0 (12):\penalty0
  124110, 2019.

\bibitem[Farrell and Oden(2014)]{farrell2014calibration}
Kathryn Farrell and J~Tinsley Oden.
\newblock Calibration and validation of coarse-grained models of atomic
  systems: application to semiconductor manufacturing.
\newblock \emph{Computational Mechanics}, 54\penalty0 (1):\penalty0 3--19,
  2014.

\bibitem[Liu et~al.(2008)Liu, Shi, Daum{\'e}~III, and Voth]{liu2008bayesian}
Pu~Liu, Qiang Shi, Hal Daum{\'e}~III, and Gregory~A Voth.
\newblock A {Bayesian} statistics approach to multiscale coarse graining.
\newblock \emph{The Journal of Chemical Physics}, 129\penalty0 (21):\penalty0
  12B605, 2008.

\bibitem[Sch{\"o}berl et~al.(2017)Sch{\"o}berl, Zabaras, and
  Koutsourelakis]{schoberl2017predictive}
Markus Sch{\"o}berl, Nicholas Zabaras, and Phaedon-Stelios Koutsourelakis.
\newblock Predictive coarse-graining.
\newblock \emph{Journal of Computational Physics}, 333:\penalty0 49--77, 2017.

\bibitem[Angelikopoulos et~al.(2012)Angelikopoulos, Papadimitriou, and
  Koumoutsakos]{angelikopoulos2012bayesian}
Panagiotis Angelikopoulos, Costas Papadimitriou, and Petros Koumoutsakos.
\newblock Bayesian uncertainty quantification and propagation in molecular
  dynamics simulations: a high performance computing framework.
\newblock \emph{The Journal of Chemical Physics}, 137\penalty0 (14):\penalty0
  144103, 2012.

\bibitem[Rixner and Koutsourelakis(2020)]{rixner2020probabilistic}
Maximilian Rixner and Phaedon-Stelios Koutsourelakis.
\newblock A probabilistic generative model for semi-supervised training of
  coarse-grained surrogates and enforcing physical constraints through virtual
  observables.
\newblock \emph{arXiv preprint arXiv:2006.01789}, 2020.

\bibitem[Xiao et~al.(2019)Xiao, Hu, Li, Attarian, Bj{\"o}rk, and
  Lendasse]{xiao2019machine}
Shaoping Xiao, Renjie Hu, Zhen Li, Siamak Attarian, Kaj-Mikael Bj{\"o}rk, and
  Amaury Lendasse.
\newblock A machine-learning-enhanced hierarchical multiscale method for
  bridging from molecular dynamics to continua.
\newblock \emph{Neural Computing and Applications}, pages 1--15, 2019.

\bibitem[Wang et~al.(2019)Wang, Olsson, Wehmeyer, P{\'e}rez, Charron,
  De~Fabritiis, No{\'e}, and Clementi]{wang2019machine}
Jiang Wang, Simon Olsson, Christoph Wehmeyer, Adri{\`a} P{\'e}rez, Nicholas~E
  Charron, Gianni De~Fabritiis, Frank No{\'e}, and Cecilia Clementi.
\newblock Machine learning of coarse-grained molecular dynamics force fields.
\newblock \emph{ACS Central Science}, 5\penalty0 (5):\penalty0 755--767, 2019.

\bibitem[Xing et~al.(2016)Xing, Triantafyllidis, Shah, Nair, and
  Zabaras]{xing2016manifold}
WW~Xing, Vasileios Triantafyllidis, Akeel~A Shah, PB~Nair, and Nicholas
  Zabaras.
\newblock Manifold learning for the emulation of spatial fields from
  computational models.
\newblock \emph{Journal of Computational Physics}, 326:\penalty0 666--690,
  2016.

\bibitem[Gorguluarslan and Choi(2014)]{gorguluarslan2014simulation}
Recep Gorguluarslan and Seung-Kyum Choi.
\newblock A simulation-based upscaling technique for multiscale modeling of
  engineering systems under uncertainty.
\newblock \emph{International Journal for Multiscale Computational
  Engineering}, 12\penalty0 (6), 2014.

\bibitem[Kennedy and O'Hagan(2000)]{kennedy2000predicting}
Marc~C Kennedy and Anthony O'Hagan.
\newblock Predicting the output from a complex computer code when fast
  approximations are available.
\newblock \emph{Biometrika}, 87\penalty0 (1):\penalty0 1--13, 2000.

\bibitem[Perdikaris et~al.(2017)Perdikaris, Raissi, Damianou, Lawrence, and
  Karniadakis]{perdikaris2017nonlinear}
Paris Perdikaris, Maziar Raissi, Andreas Damianou, Neil~D Lawrence, and
  George~Em Karniadakis.
\newblock Nonlinear information fusion algorithms for data-efficient
  multi-fidelity modelling.
\newblock \emph{Proceedings of the Royal Society A: Mathematical, Physical and
  Engineering Sciences}, 473\penalty0 (2198):\penalty0 20160751, 2017.

\bibitem[Lee et~al.(2020)Lee, Kooshkbaghi, Spiliotis, Siettos, and
  Kevrekidis]{lee2020coarse}
Seungjoon Lee, Mahdi Kooshkbaghi, Konstantinos Spiliotis, Constantinos~I
  Siettos, and Ioannis~G Kevrekidis.
\newblock Coarse-scale {PDEs} from fine-scale observations via machine
  learning.
\newblock \emph{Chaos: An Interdisciplinary Journal of Nonlinear Science},
  30\penalty0 (1):\penalty0 013141, 2020.

\bibitem[Wang et~al.(2004)Wang, Dong, and Shek]{wang2004bulk}
Wei-Hua Wang, Chuang Dong, and CH~Shek.
\newblock Bulk metallic glasses.
\newblock \emph{Materials Science and Engineering: R: Reports}, 44\penalty0
  (2-3):\penalty0 45--89, 2004.

\bibitem[Das et~al.(2005)Das, Tang, Kim, Theissmann, Baier, Wang, and
  Eckert]{das2005work}
Jayanta Das, Mei~Bo Tang, Ki~Buem Kim, Ralf Theissmann, Falko Baier, Wei~Hua
  Wang, and J{\"u}rgen Eckert.
\newblock ``work-hardenable'' ductile bulk metallic glass.
\newblock \emph{Physical Review Letters}, 94\penalty0 (20):\penalty0 205501,
  2005.

\bibitem[Schroers and Johnson(2004)]{schroers2004ductile}
Jan Schroers and William~L Johnson.
\newblock Ductile bulk metallic glass.
\newblock \emph{Physical Review Letters}, 93\penalty0 (25):\penalty0 255506,
  2004.

\bibitem[Schuh et~al.(2007)Schuh, Hufnagel, and Ramamurty]{schuh2007mechanical}
Christopher~A Schuh, Todd~C Hufnagel, and Upadrasta Ramamurty.
\newblock Mechanical behavior of amorphous alloys.
\newblock \emph{Acta Materialia}, 55\penalty0 (12):\penalty0 4067--4109, 2007.

\bibitem[Trexler and Thadhani(2010)]{trexler2010mechanical}
Morgana~Martin Trexler and Naresh~N Thadhani.
\newblock Mechanical properties of bulk metallic glasses.
\newblock \emph{Progress in Materials Science}, 55\penalty0 (8):\penalty0
  759--839, 2010.

\bibitem[Hufnagel et~al.(2016)Hufnagel, Schuh, and
  Falk]{hufnagel2016deformation}
Todd~C Hufnagel, Christopher~A Schuh, and Michael~L Falk.
\newblock Deformation of metallic glasses: Recent developments in theory,
  simulations, and experiments.
\newblock \emph{Acta Materialia}, 109:\penalty0 375--393, 2016.

\bibitem[Anand and Su(2007)]{anand2007constitutive}
Lallit Anand and Cheng Su.
\newblock A constitutive theory for metallic glasses at high homologous
  temperatures.
\newblock \emph{Acta Materialia}, 55\penalty0 (11):\penalty0 3735--3747, 2007.

\bibitem[Demetriou et~al.(2009)Demetriou, Johnson, and
  Samwer]{demetriou2009coarse}
Marios~D Demetriou, William~L Johnson, and Konrad Samwer.
\newblock Coarse-grained description of localized inelastic deformation in
  amorphous metals.
\newblock \emph{Applied Physics Letters}, 94\penalty0 (19):\penalty0 191905,
  2009.

\bibitem[Rycroft et~al.(2015)Rycroft, Sui, and
  Bouchbinder]{rycroft2015eulerian}
Chris~H Rycroft, Yi~Sui, and Eran Bouchbinder.
\newblock An {Eulerian} projection method for quasi-static elastoplasticity.
\newblock \emph{Journal of Computational Physics}, 300:\penalty0 136--166,
  2015.

\bibitem[Boffi and Rycroft(2020{\natexlab{a}})]{boffi20projection}
Nicholas~M. Boffi and Chris~H. Rycroft.
\newblock Parallel three-dimensional simulations of quasi-static elastoplastic
  solids.
\newblock \emph{Computer Physics Communications}, 257:\penalty0 107254,
  2020{\natexlab{a}}.
\newblock \doi{10.1016/j.cpc.2020.107254}.

\bibitem[Bouchbinder and Langer(2009)]{bouchbinder2009nonequilibrium}
Eran Bouchbinder and JS~Langer.
\newblock Nonequilibrium thermodynamics of driven amorphous materials. {III.}
  shear-transformation-zone plasticity.
\newblock \emph{Physical Review E}, 80\penalty0 (3):\penalty0 031133, 2009.

\bibitem[Manning and Liu(2011)]{manning2011vibrational}
M~Lisa Manning and Andrea~J Liu.
\newblock Vibrational modes identify soft spots in a sheared disordered
  packing.
\newblock \emph{Physical Review Letters}, 107\penalty0 (10):\penalty0 108302,
  2011.

\bibitem[Patinet et~al.(2016)Patinet, Vandembroucq, and
  Falk]{patinet2016connecting}
Sylvain Patinet, Damien Vandembroucq, and Michael~L Falk.
\newblock Connecting local yield stresses with plastic activity in amorphous
  solids.
\newblock \emph{Physical Review Letters}, 117\penalty0 (4):\penalty0 045501,
  2016.

\bibitem[Cubuk et~al.(2017)Cubuk, Ivancic, Schoenholz, Strickland, Basu,
  Davidson, Fontaine, Hor, Huang, Jiang, et~al.]{cubuk2017structure}
Ekin~Dogus Cubuk, RJS Ivancic, Samuel~S Schoenholz, DJ~Strickland, Anindita
  Basu, ZS~Davidson, Julien Fontaine, Jyo~Lyn Hor, Y-R Huang, Y~Jiang, et~al.
\newblock Structure-property relationships from universal signatures of
  plasticity in disordered solids.
\newblock \emph{Science}, 358\penalty0 (6366):\penalty0 1033--1037, 2017.

\bibitem[Richard et~al.(2020)Richard, Ozawa, Patinet, Stanifer, Shang, Ridout,
  Xu, Zhang, Morse, Barrat, et~al.]{richard2020predicting}
D~Richard, Misaki Ozawa, S~Patinet, E~Stanifer, B~Shang, SA~Ridout, B~Xu,
  G~Zhang, PK~Morse, J-L Barrat, et~al.
\newblock Predicting plasticity in disordered solids from structural
  indicators.
\newblock \emph{Physical Review Materials}, 4\penalty0 (11):\penalty0 113609,
  2020.

\bibitem[Langer(2004)]{langer2004dynamics}
JS~Langer.
\newblock Dynamics of shear-transformation zones in amorphous plasticity:
  Formulation in terms of an effective disorder temperature.
\newblock \emph{Physical Review E}, 70\penalty0 (4):\penalty0 041502, 2004.

\bibitem[Shi et~al.(2007)Shi, Katz, Li, and Falk]{shi2007evaluation}
Yunfeng Shi, Michael~B Katz, Hui Li, and Michael~L Falk.
\newblock Evaluation of the disorder temperature and free-volume formalisms via
  simulations of shear banding in amorphous solids.
\newblock \emph{Physical review letters}, 98\penalty0 (18):\penalty0 185505,
  2007.

\bibitem[Falk and Langer(2011)]{falk2011deformation}
Michael~L Falk and James~S Langer.
\newblock Deformation and failure of amorphous, solidlike materials.
\newblock \emph{Annu. Rev. Condens. Matter Phys.}, 2\penalty0 (1):\penalty0
  353--373, 2011.

\bibitem[Turaga et~al.(2008)Turaga, Veeraraghavan, and
  Chellappa]{turaga2008statistical}
Pavan Turaga, Ashok Veeraraghavan, and Rama Chellappa.
\newblock Statistical analysis on {Stiefel} and {Grassmann} manifolds with
  applications in computer vision.
\newblock In \emph{2008 IEEE Conference on Computer Vision and Pattern
  Recognition}, pages 1--8. IEEE, 2008.

\bibitem[Breger et~al.(2020)Breger, Orlando, Harar, D{\"o}rfler, Klimscha,
  Grechenig, Gerendas, Schmidt-Erfurth, and Ehler]{breger2020orthogonal}
Anna Breger, Jose~Ignacio Orlando, Pavol Harar, Monika D{\"o}rfler, Sophie
  Klimscha, Christoph Grechenig, Bianca~S Gerendas, Ursula Schmidt-Erfurth, and
  Martin Ehler.
\newblock On orthogonal projections for dimension reduction and applications in
  augmented target loss functions for learning problems.
\newblock \emph{Journal of Mathematical Imaging and Vision}, 62\penalty0
  (3):\penalty0 376--394, 2020.

\bibitem[Absil et~al.(2004)Absil, Mahony, and Sepulchre]{absil2004riemannian}
P-A Absil, Robert Mahony, and Rodolphe Sepulchre.
\newblock Riemannian geometry of {Grassmann} manifolds with a view on
  algorithmic computation.
\newblock \emph{Acta Applicandae Mathematica}, 80\penalty0 (2):\penalty0
  199--220, 2004.

\bibitem[Zhang et~al.(2018{\natexlab{c}})Zhang, Zhu, Heath~Jr, and
  Huang]{zhang2018grassmannian}
Jiayao Zhang, Guangxu Zhu, Robert~W Heath~Jr, and Kaibin Huang.
\newblock Grassmannian learning: Embedding geometry awareness in shallow and
  deep learning.
\newblock \emph{arXiv preprint arXiv:1808.02229}, 2018{\natexlab{c}}.

\bibitem[Ye and Lim(2016)]{ye2016schubert}
Ke~Ye and Lek-Heng Lim.
\newblock Schubert varieties and distances between subspaces of different
  dimensions.
\newblock \emph{SIAM Journal on Matrix Analysis and Applications}, 37\penalty0
  (3):\penalty0 1176--1197, 2016.

\bibitem[Jones et~al.(1998)Jones, Schonlau, and Welch]{jones1998efficient}
Donald~R Jones, Matthias Schonlau, and William~J Welch.
\newblock Efficient global optimization of expensive black-box functions.
\newblock \emph{Journal of Global Optimization}, 13\penalty0 (4):\penalty0
  455--492, 1998.

\bibitem[Williams and Rasmussen(2006)]{williams2006gaussian}
Christopher~KI Williams and Carl~Edward Rasmussen.
\newblock \emph{Gaussian processes for machine learning}, volume~2.
\newblock MIT press Cambridge, MA, 2006.

\bibitem[Rasmussen(2003)]{rasmussen2003gaussian}
Carl~Edward Rasmussen.
\newblock Gaussian processes in machine learning.
\newblock In \emph{Summer School on Machine Learning}, pages 63--71. Springer,
  2003.

\bibitem[Gramacy(2020)]{gramacy2020surrogates}
Robert~B Gramacy.
\newblock \emph{Surrogates: Gaussian Process Modeling, Design, and Optimization
  for the Applied Sciences}.
\newblock CRC Press, 2020.

\bibitem[Aggarwal et~al.(2001)Aggarwal, Hinneburg, and
  Keim]{aggarwal2001surprising}
Charu~C Aggarwal, Alexander Hinneburg, and Daniel~A Keim.
\newblock On the surprising behavior of distance metrics in high dimensional
  space.
\newblock In \emph{International conference on database theory}, pages
  420--434. Springer, 2001.

\bibitem[Hamm and Lee(2008)]{hamm2008grassmann}
Jihun Hamm and Daniel~D Lee.
\newblock Grassmann discriminant analysis: a unifying view on subspace-based
  learning.
\newblock In \emph{Proceedings of the 25th international conference on Machine
  learning}, pages 376--383, 2008.

\bibitem[Plimpton(1995)]{plimpton1995fast}
Steve Plimpton.
\newblock Fast parallel algorithms for short-range molecular dynamics.
\newblock \emph{Journal of Computational Physics}, 117\penalty0 (1):\penalty0
  1--19, 1995.

\bibitem[Cheng et~al.(2009)Cheng, Ma, and Sheng]{cheng2009atomic}
YQ~Cheng, E~Ma, and HW~Sheng.
\newblock Atomic level structure in multicomponent bulk metallic glass.
\newblock \emph{Physical Review Letters}, 102\penalty0 (24):\penalty0 245501,
  2009.

\bibitem[Sheng et~al.(2011)Sheng, Kramer, Cadien, Fujita, and
  Chen]{sheng2011highly}
HW~Sheng, MJ~Kramer, A~Cadien, T~Fujita, and MW~Chen.
\newblock Highly optimized embedded-atom-method potentials for fourteen fcc
  metals.
\newblock \emph{Physical Review B}, 83\penalty0 (13):\penalty0 134118, 2011.

\bibitem[Evans and Morriss(1984)]{evans1984nonlinear}
Denis~J Evans and GP~Morriss.
\newblock Nonlinear-response theory for steady planar {Couette} flow.
\newblock \emph{Physical Review A}, 30\penalty0 (3):\penalty0 1528, 1984.

\bibitem[Alix-Williams and Falk(2018)]{alix2018shear}
Darius~D Alix-Williams and Michael~L Falk.
\newblock Shear band broadening in simulated glasses.
\newblock \emph{Physical Review E}, 98\penalty0 (5):\penalty0 053002, 2018.

\bibitem[Rycroft and Bouchbinder(2012)]{rycroft2012fracture}
Chris~H Rycroft and Eran Bouchbinder.
\newblock Fracture toughness of metallic glasses: Annealing-induced
  embrittlement.
\newblock \emph{Physical Review Letters}, 109\penalty0 (19):\penalty0 194301,
  2012.

\bibitem[Vasoya et~al.(2016)Vasoya, Rycroft, and Bouchbinder]{vasoya16fracture}
Manish Vasoya, Chris~H. Rycroft, and Eran Bouchbinder.
\newblock Notch fracture toughness of glasses: Dependence on rate, age, and
  geometry.
\newblock \emph{Phys. Rev. Applied}, 6:\penalty0 024008, Aug 2016.
\newblock \doi{10.1103/PhysRevApplied.6.024008}.

\bibitem[Boffi and Rycroft(2020{\natexlab{b}})]{boffi20transform}
Nicholas~M. Boffi and Chris~H. Rycroft.
\newblock Coordinate transformation methodology for simulating quasistatic
  elastoplastic solids.
\newblock \emph{Phys. Rev. E}, 101:\penalty0 053304, May 2020{\natexlab{b}}.
\newblock \doi{10.1103/PhysRevE.101.053304}.

\bibitem[Falk and Langer(1998)]{falk1998dynamics}
Michael~L Falk and James~S Langer.
\newblock Dynamics of viscoplastic deformation in amorphous solids.
\newblock \emph{Physical Review E}, 57\penalty0 (6):\penalty0 7192, 1998.

\bibitem[Vesselinov et~al.(2019)Vesselinov, Mudunuru, Karra, O'Malley, and
  Alexandrov]{vesselinov2019unsupervised}
Velimir~V Vesselinov, Maruti~Kumar Mudunuru, Satish Karra, Dan O'Malley, and
  Boian~S Alexandrov.
\newblock Unsupervised machine learning based on non-negative tensor
  factorization for analyzing reactive-mixing.
\newblock \emph{Journal of Computational Physics}, 395:\penalty0 85--104, 2019.

\end{thebibliography}

\end{document}